\documentclass[
reprint,
superscriptaddress,
 amsmath,
 amssymb,
 aps,
]{revtex4-1}

\usepackage{tabularx}
\usepackage[normalem]{ulem} 
\usepackage{multirow} 
\usepackage{cancel}
\usepackage{braket}
\usepackage{graphicx}
\usepackage{dcolumn}
\usepackage{bm}
\usepackage{nicefrac}
\usepackage[mathscr]{euscript}
\usepackage{soul} 

\newcommand{\abs}[1]{{\left\lvert #1 \right\rvert}}

\usepackage[usenames,dvipsnames]{xcolor}
\newcommand{\red}       [1]   {\textcolor{red}       {#1}} 
\newcommand{\blue}      [1]   {\textcolor{blue}      {#1}} 

\begin{document}


\newcommand{\Vswarm}{\mathscr{V}} 
\newcommand{\newL}{\marginpar{ \red{\footnotesize $\leftarrow$NEW}}} 
\newcommand{\new}{\marginpar{  \red{\footnotesize NEW$\to$}}}

\newcommand{\nm}{\mathrm{nm}}
\newcommand{\calF}{\mathcal{F}}
\newcommand{\calN}{{\scriptscriptstyle \mathcal{N}} }
\newcommand{\reta}{\red{\eta}}
\newcommand{\blues}{\blue{s}}
\newcommand{\calKconf}{\calK^{\mathrm{conf}}}
\newcommand{\calKedge}{\calK^{\mathrm{edge}}}

\newcommand{\AEdge}{A^{\text{edge} }   }
\newcommand{\Aconf}{A^{\text{conf}}   }
\newcommand{\Cconf}{C^{\text{conf}}   }
\newcommand{\Cedg}{C^{\text{edge}}   }
\newcommand{\K}{\mathbf{K}}
\newcommand{\PHIAKp}{ \Phi_A^{\K'}(y) }
\newcommand{\PHIAK}{ \Phi_A^{\K}(y) }
\newcommand{\PHIBKp}{ \Phi_B^{\K'}(y) }
\newcommand{\PHIBK}{ \Phi_B^{\K}(y) }
\newcommand{\Weff}{W}  

\newcommand{\ra}{{\mathrm{a}}}
\newcommand{\rb}{{\mathrm{b}}}
\newcommand{\rc}{{\mathrm{c}}}
\newcommand{\rd}{{\mathrm{d}}}

\newcommand{\rmi}{\mathrm{\,i}}
\newcommand{\rmj}{\mathrm{\,j}}

\newcommand{\calK}{\mathcal{K}}

\newcommand{\kappax}{\kappa_x}
\newcommand{\kedge}{\mathcal{K}^{\text{edge}}}
\newcommand{\kconf}{\mathcal{K}^{\text{conf}}}

\newcommand{\signConfined}{\zeta^{\text{conf}}   }
\newcommand{\signEdge}{\zeta^{\text{edge}}}

\newcommand{\rmR}{\mathrm{R}}
\newcommand{\rmC}{\mathrm{C}}
 \newcommand{\calX}{\mathcal{X}}

\newcommand{\xhat}{\hat{\bm{x}}}
\newcommand{\yhat}{\hat{\bm{y}}}


\preprint{APS/123-QED}

\title{Coherent control of current injection in zigzag graphene nanoribbons}
\author{Cuauht\'emoc Salazar}
\affiliation{Department of Physics and Institute of Optical Sciences, University of Toronto,\\
60 St. George Street, Toronto, Ontario, Canada, M5S 1A7}
\author{J. L. Cheng}
\affiliation{Department of Physics and Institute of Optical Sciences, University of Toronto,\\
60 St. George Street, Toronto, Ontario, Canada, M5S 1A7}
\affiliation{Brussels Photonics Team (B-PHOT), Department of Applied Physics and
Photonics (IR-TONA), Vrije Universiteit Brussel, Pleinlaan 2, 1050
Brussel, Belgium}
\author{J. E. Sipe}
\affiliation{Department of Physics and Institute of Optical Sciences, University of Toronto,\\
60 St. George Street, Toronto, Ontario, Canada, M5S 1A7}
\date{\today}
\begin{abstract} 
  We present Fermi's  golden rule calculations of the optical carrier
  injection and the coherent control of current injection in graphene
  nanoribbons with zigzag geometry, using an envelope function
  approach. This system possesses strongly localized states (flat
  bands) with a large joint density of states at low photon energies;
  for ribbons with widths above a few tens of nanometers, this system also
  posses large number of (non-flat) states with maxima and minima
  close to the Fermi level. Consequently, even with small dopings the
  occupation of these localized states can be significantly
  altered. In this work, we calculate the relevant quantities for
  coherent control at different chemical potentials, showing the
  sensitivity of this system to the occupation of the edge states. We
  consider coherent control scenarios arising from the interference of
  one-photon absorption at $2\hbar\omega$ with two-photon absorption
  at $\hbar\omega$, and those arising from the interference of
  one-photon absorption at $\hbar\omega$ with stimulated electronic
  Raman scattering (virtual absorption at $2\hbar\omega$ followed by
  emission at $\hbar\omega$).  Although at large photon energies these
  processes follow an energy-dependence similar to that of 2D
  graphene, the zigzag nanoribbons exhibit a richer structure at low
  photon energies, arising from divergences of the joint density of
  states and from resonant absorption processes, which can be strongly
  modified by doping. As a figure of merit for the injected carrier
  currents, we calculate the resulting swarm velocities. Finally, we
  provide estimates for the limits of validity of our model.

%
\end{abstract}



\maketitle

\section{Introduction\label{sec:intro}}
The electronic properties of low-dimensional materials depend strongly
on the size and geometry of the system
\cite{Ogawa_optics_low_dimensions, Nakada_edges_early}. For instance,
the bandstructure of a monolayer and a stripe of graphene are
significantly different. A stripe of graphene is usually referred as a
graphene nanoribbon, where the boundaries impose novel conditions on
the wavefunctions; for a zigzag graphene nanoribbon (ZGNR), the
wavefunction vanishes on a single sublattice, A or B, at each edge. As
shown earlier \cite{Nakada_edges_early, BreyFertig, marconciniKDP}, in
ZGNR, there are \textit{confined states} that extend across the width
of the system, incorporating states from both sublattices. There is
also another class of states strongly localized at each edge, which
incorporate states from either one or the other sublattice; these
states are known as \textit{edge states}. 
Although confined states are also found in other types of ribbons,
such as armchair, the edge states are present only for zigzag
ribbons. %
These edge and confined states provide many of the novel
  characteristics seen in ZGNR. %
  Moreover, the energy of these states can be easily tuned by changing
  the ribbon width, applying external fields, and functionalizing the
  system~\cite{Lee_functionalization,daRocha_functionalization}. Since for an undoped ZGNR the Fermi level coincides with the
  flat part of the edge states, then tuning the doping level allows to
  easily control the contribution of the edge states. %
  Given that a 2D graphene sheet lacks of these localized states, a ZGNR
  offers the advantage of having optical responses that are easily
  tuneable. %
Over the last years, a number of studies have reported the special
properties of these localized states
\cite{Nakada_edges_early,BreyFertig, marconciniKDP,
  Yang_Cohen_Louie_2008,
  Kunstmann_Stability, Bellec_edges_artificial_graphene,
  Delplace_edges_zak_phase} and recent investigations have described
more novel properties and applications ~\cite{Yazyev_applications,
  luck2015, Fujita_EELS_2015, pelc-brey, gonsalbez-spin-filtered,
  Stegmann}.
At zero energy they have an important role in the electronic transport
properties of both clean and disordered ZGNR, as Luck \textit{et
  al}.~\cite{luck2015} (and references therein) have recently shown
using a tight-binding formalism with a transfer-matrix approach.
A detailed review of these localized states in graphene-like systems can be found
in Lado \textit{et al.}  \cite{lado2015}.
The optical properties of ZGNR and graphene nano-flakes have been
studied from a number of perspectives \cite{Hsu-Reich_selection_rules,
  Sasaki_optical_transitions, Yamamoto2006_TB_optics_flakes,
  Berahman_optics_chiral, zhu_optical_conduc, Yang_Cohen_Louie_2008,
  Prezzi_2008_optics, Fujita_EELS_2015}, always
showing the strong influence of the edge states in the dielectric
function.
First-principles studies of functionalization in graphene ribbons have
shown~\cite{Lee_functionalization} that the low-energy $\pi$ electrons
at the edges of the ZGNR lead to higher binding energies as compared
with ribbons of different shape edges. Similar studies
indicate~\cite{daRocha_functionalization} that the optical response of
functionalized ZGNR depends strongly on the size, shape and location
of the deposited molecule, suggesting functionalization as an
effective way of fine-tuning the electronic and optical properties of
ZGNR.

In this work, we investigate the optical injection of carriers and
currents in graphene nanoribbons by means of coherent light fields at
$\omega$ and $2\omega$. 
In general, for arbitrary beams, this technique is referred as
\textit{coherent current control}. It is based on the fundamental
feature that if the quantum evolution of a system can proceed via
several pathways, then the interference between such pathways can play
a determining role in the final state of the system
\cite{ManykinAfanas,Manykin}. In a semiconductor, it is possible to control
the injection of
carriers \cite{ccontrolDrielSipe2001,sun-norris2010,Rioux2011,fregoso},
spins, electrical current~\cite{kiran2012}, spin
current~\cite{Muniz_SpinTopos}, and even valley
current~\cite{Muniz_ValleyCurrent15}, using phase-dependent
perturbations, usually involving coherent beams or pulses of light.
In a one-color scheme, the interference is between transition
amplitudes associated with different
polarizations~\cite{ccontrolDrielSipe2001}.  Although carrier
injection can be achieved with one-color excitation, current injection
cannot. This is due to symmetry considerations, since one-color
current injection is characterized by a third-rank tensor, hence it is
only allowed in systems that lack inversion
symmetry~\cite{ccontrolDrielSipe2001}. Due to the inversion symmetry
in zigzag graphene ribbons, the one-color coherent control process is
forbidden. In a two-color scheme, the interference is between pathways
related to photon absorption processes arising from different phase
related beams, one at $\omega$ and the other at $2\omega$. In
  this case, current injection is characterized by a fourth-rank
  tensor, hence it is nonzero for a ZGNR. In both schemes, the
different pathways connect the same initial and final states.  Here
our focus is on two-color current injection, and we consider two
classes of processes: 
the first class arises from the interference of one-photon absorption
at $2\hbar\omega$ with two-photon absorption at $\hbar\omega$, and the
second class arises from the interference of one-photon absorption at
$\hbar\omega$ with stimulated electronic Raman scattering at
$\hbar\omega$.
In general, coherent control injection allows for the placement of
electrons and holes in different bands and portions of the Brillouin
Zone as $\omega$ is varied. Thus, as we will show, the current
injection is very sensitive to the presence of both confined and edge
states. 
In line with plausible experiments, we consider nanoribbons with a
width on the order of 20~nanometers, which leads to unit cells
containing a few hundreds of atoms. For this reason, we employ an
envelope function strategy to calculate the relevant energies and
velocity matrix elements; the rest of the calculation follows a
conventional Fermi's golden rule approach to calculate the absorption
coefficients.

The article is organized as follows. In Sec.~\ref{sec:model}, we
describe the model Hamiltonian employed to describe the wavefunctions,
the resulting bandstructure, and the selection rules for the velocity
matrix elements. In Sec.~\ref{sec:CC}, we describe the different
carrier injection and current injection coefficients, including the
conventional and Raman contributions. In Sec.~\ref{sec:doping}, we
revisit these calculations, but for a $p$-doped system. This allows us to
show the significant change in the signals that can be accomplished by
altering the occupation of the edge states. In
Sec.~\ref{sec:estimates}, we provide an estimate of the limits of validity of
the model employed in this work. Finally, in
Sec.~\ref{sec:discussion}, we present our final discussions and
conclusions.
\section{Theoretical Model \label{sec:model}}
\subsection{Model Hamiltonian}  
\begin{figure}[ht]
  \centering
  \includegraphics[scale=1]{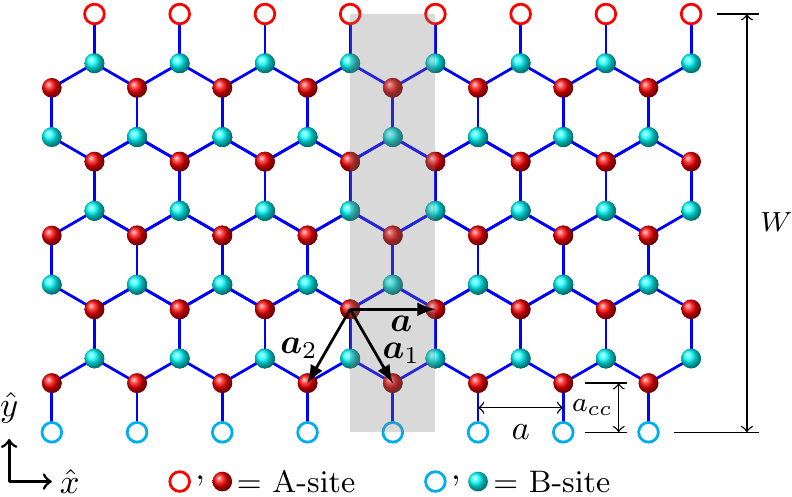}
  \caption{(Color online) Illustration of the lattice structure of a
    zigzag graphene nanoribbon extended along $\xhat$ and confined
    along $\yhat$. Passivation atoms and carbon atoms are represented
    by unfilled and filled circles, respectively; A (B) sites are
    colored red (cyan) and the unit cell is represented in grey. }
  \label{fig:structure}
\end{figure}

A zigzag graphene nanoribbon (ZGNR) is a strip of monolayer graphene
\cite{CastroNetoReview2009, DasSarmaReview2011} that has been cut such
that the edges along its length have a zigzag shape, as shown in
Fig.~\ref{fig:structure}. We take the ribbon to lie in the ($xy$)
plane, with $\xhat$ as the longitudinal direction along which the
ribbon extends over all space; $\yhat$ then identifies the direction
across the ribbon, along which the electron states are confined.

We assume passivated carbon atoms at the longitudinal boundaries, as
if hydrogen atoms were adsorbed~\cite{marconciniKDP,Fujita_EELS_2015};
this allows the passivation of any dangling edge states and the
neutralization of  the spin moments at the edges~\cite{Fujita_EELS_2015}. We
take $W=a\sqrt{3}\,(2N+2)/6$ as the effective width, where $N$ is the
total number of atoms in the unit cell,
$a=a_{\mathrm{cc}}\sqrt{3}=0.246~\mathrm{nm}$ is the graphene lattice
constant, and $a_{\mathrm{cc}}$ is the carbon-carbon distance (see
Fig.~\ref{fig:structure}).
The edge at $y=a/\sqrt{3}$ is formed by A-atoms, while the edge at
$y=W-a/\sqrt{3}$ is formed by B-atoms.  The lattice vector is
  $\bm{a}=a \xhat$ and the atomic sites are set in terms of the
  graphene lattice vectors, $\bm{a}_1=( \xhat -\sqrt{3}\yhat )\, a/2 $
  and $\bm{a}_2= (-\xhat -\sqrt{3}\yhat)\, a/2$. The Dirac points of
  monolayer graphene are projected~\cite{marconciniKDP} into the one-dimensional
  Brillouin zone of the ZGNR, $[-\frac{\pi}{a}, \frac{\pi}{a})$, as
  $\K=\frac{2\pi}{3a}$ and $\K'=-\frac{2\pi}{3a}$. 
We express the total wavefunctions as linear combinations of
  atomic orbitals $\varphi$ that are centered at atomic sites A and B,
\begin{align}
\label{eq:total_wf}
  \Psi(\bm{r}) &= %
\sum\limits_{\bm{R}_A} \psi_A(\bm{R}_A) \varphi(\bm{r}-\bm{R}_A) \notag\\%
& + %
\sum\limits_{\bm{R}_B} \psi_B(\bm{R}_B) \varphi(\bm{r}-\bm{R}_B). %
\end{align}
Then, following Marconcini and Macucci \cite{marconciniKDP}, we employ
the semi-empirical $\bm{k}\cdot\bm{p}$ method to describe
$\Psi(\bm{r})$ with a smooth envelope function approach. The
coefficients $\psi_A$ and $\psi_B$ in Eq.~\eqref{eq:total_wf} can be
written as
\begin{subequations}
 \label{eq:total_wf_coeffs}
\begin{align}
  \psi_A(\bm{r}) &= e^{i \K\cdot\bm{r}} F_A^\K(\bm{r}) + e^{i \K'\cdot\bm{r}} F_A^{\K'}(\bm{r}),
\\
  \psi_B(\bm{r}) &= -e^{i \K\cdot\bm{r}} F_B^\K(\bm{r}) + e^{i \K'\cdot\bm{r}} F_B^{\K'}(\bm{r}), 
\end{align}
\end{subequations}
where the $F_{A(B)}^{\bm{K} (\bm{K}')} (\mathbf{r})$ are the envelope
function components associated with the $\bm{K} (\bm{K}')$ Dirac point
and the orbital at atom A(B)\footnote{The graphene's honeycomb lattice
  is composed by two distinct triangular lattices, A and B. On each
  sub-lattice all atoms are equivalent.}. In writing
Eq.~\eqref{eq:total_wf_coeffs} we have replaced
$\psi_i(\bm{R}_i)\to\psi_i(\bm{r})$ for $i=\{A, B\}$, on the basis of
two assumptions. First, we assume that atomic orbitals are strongly localized
at their corresponding atom, and second, we assume that the envelope
functions are slow-varying functions of $\bm{r}$ near the $\K$ ($\K'$)
Dirac point.  These envelope functions satisfy the Dirac equation,
\begin{eqnarray}
  \label{eq:dirac}
  \begin{bmatrix}
    0 & -i\partial_x-\partial_y  & 0  & 0 \\
-i \partial_x +\partial_y & 0 & 0 &  0 \\
  0                                  & 0 & 0& -i\partial_x+\partial_y \\
0 & 0  & -i\partial_x -\partial_y & 0
  \end{bmatrix}
\nonumber\\
\times
  \begin{bmatrix}
F_A^{\mathbf{K}} (\mathbf{r})\\
F_B^{\mathbf{K}} (\mathbf{r})\\
F_A^{\mathbf{K}'} (\mathbf{r})\\
F_B^{\mathbf{K}'} (\mathbf{r})\\
  \end{bmatrix}
= \frac{E}{\gamma}\;
  \begin{bmatrix}
F_A^{\mathbf{K}}(\mathbf{r})\\
F_B^{\mathbf{K}}(\mathbf{r})\\
F_A^{\mathbf{K}'}(\mathbf{r})\\
F_B^{\mathbf{K}'}(\mathbf{r})\\
  \end{bmatrix},
\qquad
\end{eqnarray}
where $\gamma=(\sqrt{3}/2)\,t a$, $t =2.70~\mathrm{eV}$ is the
nearest-neighbor hopping parameter and $v_F=\gamma\hbar^{-1}$ is the
graphene Fermi velocity. 
Because of the translational symmetry along $\xhat$, each envelope
function can be factorized as the product of a propagating plane wave
along the length direction ($\xhat$), and a function confined along
the width direction ($\yhat$),
\begin{eqnarray}
  \label{eq:envelopeK}
\bm{F^K}(\bm{r})&\equiv
  \begin{bmatrix}
    F_A^{\mathbf{K}} (\mathbf{r})\\
    F_B^{\mathbf{K}} (\mathbf{r}) 
  \end{bmatrix}
=
e^{i  \kappa_x x}\;
  \begin{bmatrix}
\Phi_A^{\mathbf{K}} (y)\\
\Phi_B^{\mathbf{K}} (y)
  \end{bmatrix},
\\
\label{eq:envelopeKp}
\bm{F^{K'}}(\bm{r})&\equiv
  \begin{bmatrix}
    F_A^{\mathbf{K'}} (\mathbf{r})\\
    F_B^{\mathbf{K'}} (\mathbf{r})
  \end{bmatrix}
=
e^{i  \kappa_x' x}\;
  \begin{bmatrix}
\Phi_A^{\mathbf{K'}} (y)\\
\Phi_B^{\mathbf{K'}} (y)
  \end{bmatrix},
\end{eqnarray}
where $\kappax$ ($\kappax'$) is the wavevector along the length of the
ribbon, measured from the Dirac point $\bm{K}$ ($\bm{K}'$). 
The passivation of the carbon atoms at the edges terminates the
$\pi$ orbitals thereat, thus it is reasonable to assume
that the full wavefunction vanishes at the lattice sites located at
the effective edges.  This leads to the following boundary conditions for the confined
part of the envelope functions~\cite{marconciniKDP},
\begin{subequations}
  \label{eq:bconditions}
\begin{align}
  \label{eq:bconditionsK}
  \Phi_B^{\bm{K}} (y=0)  &=0,  &\Phi_A^{\bm{K}} (y=\Weff) & =0,\\
  \label{eq:bconditionsKp}
    \Phi_B^{\bm{K'}} (y=0) &=0,  &\Phi_A^{\bm{K'}} (y=\Weff) & =0.
\end{align}
\end{subequations}
These boundary conditions and the block diagonal form of the matrix in
Eq.~\eqref{eq:dirac} cause the envelope functions at $\bm{K}$ to be
uncoupled from their counterparts at $\bm{K}'$; therefore they can be
studied separately. %
With the use of Eq.~\eqref{eq:envelopeK}, the Dirac equation for the
$\bm{K}$ valley is 
\begin{eqnarray}
  \label{eq:confinedK}
\gamma
   \begin{bmatrix}
0 & \kappa_x - \partial_y \\
\kappa_x + \partial_y & 0
   \end{bmatrix} 
  \begin{bmatrix}
\Phi_A^{\mathbf{K}} (y)\\
\Phi_B^{\mathbf{K}} (y)
  \end{bmatrix}
= E
  \begin{bmatrix}
\Phi_A^{\mathbf{K}} (y)\\
\Phi_B^{\mathbf{K}} (y)
  \end{bmatrix}.
 \end{eqnarray}
The solutions of Eq.~\eqref{eq:confinedK} are of the form \cite{marconciniKDP},
\begin{align}
  \label{eq:21}
  \Phi_A^{\mathbf{K}} (y) &= \frac{\gamma}{E} \Big[ (\kappa_x-\calK) A
                            e^{\calK y} + (\kappa_x+\calK) B e^{-\calK y} \Big], \\
  \Phi_B^{\mathbf{K}} (y) &= A e^{\calK y} + B e^{-\calK y},
\end{align}
where $\calK=\sqrt{\kappa_x^2-(E/\gamma)^2}$. Under the boundary
conditions (Eq.~\eqref{eq:bconditionsK}), this leads to a relation between the
transverse ($\calK$) and the longitudinal ($\kappax$) wavenumbers,
\begin{eqnarray}
  \label{eq:dispersion}
  e^{-2\calK \Weff} = \frac{\kappax-\calK}{\kappax+\calK},
\end{eqnarray}
which shows that they are coupled for ZGNR. 
If $\calK$ is taken to be real, then Eq.~\eqref{eq:dispersion} reduces to
\begin{eqnarray}
  \label{eq:299}
  \kappax = \calK \; \coth\left( \Weff \calK \right),
\end{eqnarray}
and without loss of generality we assume $\calK$ to be
positive. Equation~\eqref{eq:299} supports two eigensolutions for
$\kappax>\Weff^{-1}$, which we label as $n=1$ for positive energies
and $n=-1$ for negative energies; both correspond to states strongly
confined at the edges, henceforth referred as \textit{edge
  states}~\cite{marconciniKDP},
\begin{eqnarray}
  \label{eq:4A}
  \PHIAK &=&\frac{-2}{\sqrt{{L}}} \AEdge \; \signEdge_n \; \sinh\; \left[\kedge  (\Weff-y)\right],\;   \\
  \label{eq:4B}
  \PHIBK&=& \frac{2}{\sqrt{{L}}}  \AEdge \sinh \left[ \kedge y  \right], \\
  \signEdge_n &=& n, \quad\text{for~} n=\pm1, \label{eq:sign_edge}
\end{eqnarray}
 where $L$ is a
normalization length along the $\xhat$ direction.
We have also set $\calK\to \kedge$, and $\AEdge$  is the usual
wavefunction normalization coefficient, 
\begin{eqnarray}
  \label{eq:3}
      \AEdge &=& \sqrt{  \frac{\kedge / 2}   {\sinh (2 \kedge\Weff) -
      (2 \kedge\Weff) }    },
\end{eqnarray}
and the eigenenergy is 
\begin{eqnarray}
\label{eq:eigenEnergiesEdge}
  E_n^{\mathrm{edge}} = n\; \gamma \sqrt{\kappax^2-(\calK^{\mathrm{edge}})^2}.
\end{eqnarray}
Conversely, if we consider solutions of Eq.~\eqref{eq:dispersion} with
$\calK$ purely imaginary, of the form $i\calK_n$ with $\calK_n$ real,
then Eq.~\eqref{eq:dispersion} reduces to
\begin{eqnarray}
  \label{eq:2}
  \kappax = \calK_n \; \cot\left( \Weff \calK_n \right),
\end{eqnarray}
where, without loss of generality,  we take $\calK_n$ to be
positive.  These solutions give states that extend over the full width
of the ribbon, and are known simply as \textit{confined states};
for these we set $\calK_n\to\calK_n^{\mathrm{conf}}$ and label
them by $n=\pm1, \pm2, \pm3, \ldots$, starting with $\pm1$ for those
 with energies closest to zero. 
These confined states  exist for any real $\kappax$, except those with
band index $n=\pm1$, which exist only for $\kappax{\leq}\Weff^{-1}$.
The dispersion relations of the confined states with band index
$n=\pm1$  connect with that of the edge states; both share the band index $n=\pm1$ %
({transition from the red to the blue traces in
  Fig.~\ref{fig:bands_K}}). %
The confined states have the form
\begin{eqnarray}
  \label{eq:5A}
    \PHIAK&=&-i \frac{2}{\sqrt{{L}}}  \Aconf_n \; \signConfined_n \; \sin \left[ \calK_n^{\mathrm{conf}} (\Weff-y) \right],\\
  \label{eq:5B}
    \PHIBK&=& \phantom{-} i \frac{2}{\sqrt{{L}}}   \Aconf_n \sin \left[  \calK_n^{\mathrm{conf}} y  \right], \\
    \signConfined_n &=&  (-1)^{n+1} sgn(n), 
\end{eqnarray}
where
\begin{eqnarray}
  \label{eq:9}
  \Aconf_n &=& \sqrt{  \frac{\kconf_n /2}   {-\sin (2 \kconf_n \Weff) +
      (2 \kconf_n \Weff) }    }, \\
\label{eq:eigenEnergiesConf}
  E_n^{\mathrm{conf}} &=& sgn(n) \, \gamma \, \sqrt{\kappax^2 +
    (\calK_n^{\mathrm{conf}})^2 }.
\end{eqnarray}
We can indicate any of the edge or confined states simply  by $|n\kappax \rangle$,
where if $|n|\ge2$ the state is confined, while if $|n|=1$ then the state
is confined for $\kappax{\leq}\Weff^{-1}$, but it is an edge state if
$\kappax>\Weff^{-1}$. 

\begin{figure}[ht]
  \centering
  \includegraphics[scale=0.7]{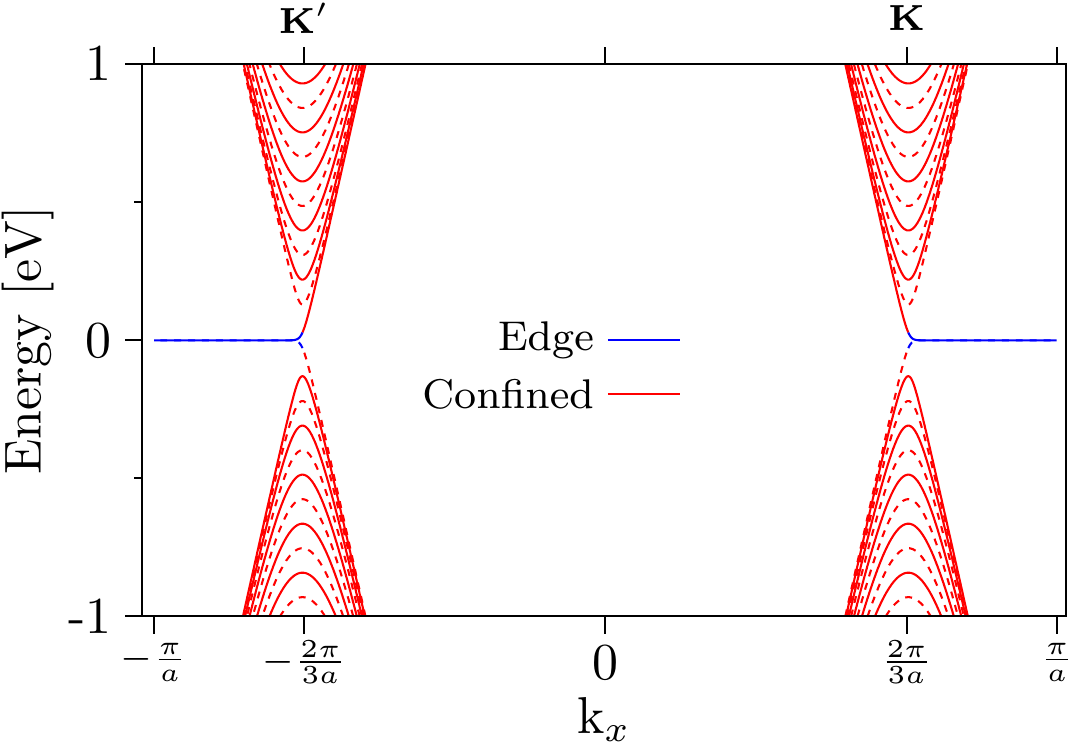}
  \caption{(Color online) Zigzag nanoribbon bandstructure with 95 zigzag lines
    (about $20$~nm width). Solid and dashed lines
    distinguish the polarity of the states.  The  confined states
    are shown in red and red-dashed lines, while the edge states are
    shown in blue and blue-dashed lines. The latter are flattened
    towards zero energy. The different polarities of these edge states
    is more distinguishable in the 
    inset given in Fig.~\ref{fig:bands_K}. 
    The horizontal axis corresponds to the total wavevectors
    $\mathrm{k}_x$, measured from the Brillouin zone center,
    cf. Fig.~\ref{fig:bands_K}. %
  \label{fig:bands_both_K}
}
\end{figure}

\begin{figure}[ht]
  \centering
  \includegraphics[scale=0.8]{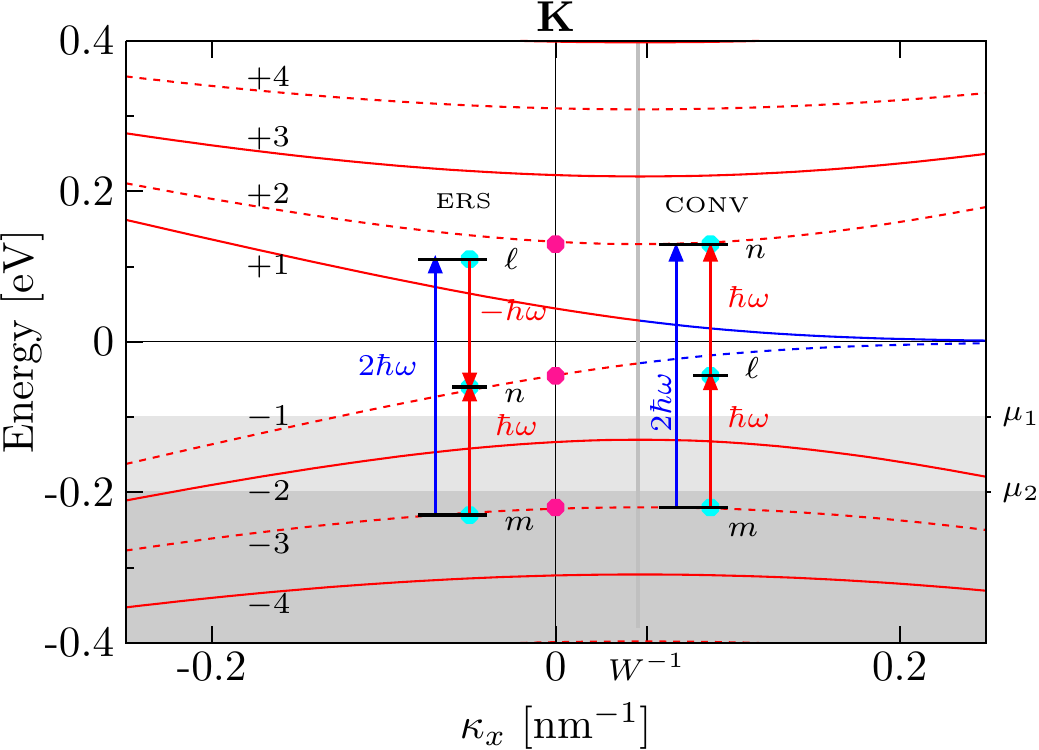}
  \caption{(Color online)  Depiction of the conventional coherent control (CC) scheme
    (set of arrows on the right) and the ERS CC (left arrows).
    Confined and edge states are shown in red and blue lines,
    respectively; solid and dashed lines
    distinguish the polarity of the states (see also Fig.~\ref{fig:bands_both_K}). The
    initial (final) state is $m$ ($n$) and $\ell$ is a virtual
    state. For $m=-3, n=2, \ell=-1$, the three purple dots along
    $\kappa_x=0$ pinpoint three states at which both the conventional
    and the ERS current injection are resonant. The upper boundaries
    of the grey areas depict Fermi levels of $\mu_1=-0.10~$eV and $\mu_2=-0.20~$eV
    ($p$-doped system).  The horizontal axis corresponds to wavevectors
    $\bm{\kappa}_x$ measured from the Dirac point $\bm{K}$,
    cf. Fig.~\ref{fig:bands_both_K}.  The vertices of the parabolic
    (confined) states occur at $\kappax=\Weff^{-1}$.
    \label{fig:bands_K}}
\end{figure}

Equations \eqref{eq:eigenEnergiesEdge} and
\eqref{eq:eigenEnergiesConf} describe the bandstructure of ZGNR, shown
in Figs.~\ref{fig:bands_both_K} and \ref{fig:bands_K}. The edge
states  are flattened towards the zero energy level for
$\kappa_x>W^{-1}$ (Fig.~\ref{fig:bands_K}), whereas the confined
states  have a parabolic structure around the Dirac points, with
an axis of symmetry at $\kappax=\Weff^{-1}$, except for the two
confined states nearest to zero energy, with band index $n=\pm1$ and
$\kappa_x\leq W^{-1}$ (Fig.~\ref{fig:bands_K}). These confined states
are associated with the Dirac cones of 2D graphene. %
Since the extrema of the confined states occur at $\kappa_x= W^{-1}$ ,
we can express the band energies at such value of $\kappa_x$ as
\begin{subequations}
  \label{eq:gap_scaling}
  \begin{align}
    E_{\pm 1}(W^{-1}) &= \pm\gamma W^{-1},\\
E_{\pm n} (W^{-1})&\approx \pm  \gamma  W^{-1}\sqrt{1+\pi^2\left(n-\nicefrac{1}{2}\right)^2},
  \end{align}
\end{subequations}
  for the edge and confined states, respectively. This indicates that
  the band gap scales as $W^{-1}$ and provides an estimate of the
  photon energy at which the absorption edge occurs with respect to
  the ribbon width $W$.
It turns out that the sign functions appearing in the expressions for
$\PHIAK$ [Eq.~\eqref{eq:4A} for edge states and Eq.~\eqref{eq:5A} for
confined states] alternate  for consecutive states, being $+1$ for
the first state above zero energy, $-1$ for the next up, and so on;
the situation is reversed for negative energies. This sign factor
plays an important role in the selection rules of the quantities we
calculate. Therefore we indicate these sign factors on the bandstructure
diagram [Figs.~\ref{fig:bands_both_K} and \ref{fig:bands_K}): a solid line indicates that the confined part of an A-site
component of the envelope function has  $\zeta_n=+1$, whereas a dashed
trace means it has $\zeta_n=-1$.

\subsection{Velocity matrix elements}
We employ the envelope functions given by %
Eq.~\eqref{eq:envelopeK} in order to calculate the velocity matrix
elements (VME) that describe the coupling between two states
$|n,\kappa_x\rangle$ and $|m,\kappa_x\rangle$ as,
\begin{eqnarray}
\label{eq:10}
\bm{v}_{nm} (\kappa_x)   &=& %
\int d\bm{r}\, \left[\bm{F^K}(\bm{r})\right]^\dagger \bm{v} \left[\bm{F^K}(\bm{r})\right],
\end{eqnarray}
where $\kappa_x$ is a wavenumber and $n$, $m$ are band indices. The
velocity operator is given by
\mbox{$\bm{v}=\left[\bm{r}, H\right]/(i\hbar)$}, which, together with
the Hamiltonian in Eq.~\eqref{eq:confinedK} for the $\K$ valley, 
\begin{eqnarray}
  \label{eq:hamilton}
H=
\gamma %
  \begin{bmatrix}
    0 & -i\partial_x-\partial_y   \\
    -i \partial_x +\partial_y & 0 
\end{bmatrix},
\end{eqnarray}
leads to
\mbox{$\bm{v} = v_F (\sigma_x, \sigma_y)$},
where $\sigma_x$ and $\sigma_y$ are
the Pauli matrices and $v_F=\gamma/\hbar$ is graphene's Fermi
velocity.
The resulting expressions are given in
Appendix~\ref{appendixVME}, Table~\ref{tab:vmetable}, and 
obey the following selection
rules:
\begin{subequations}
  \label{eq:vmeSelectionRules}
\begin{eqnarray}
v^x_{nm} (\kappa_x)  = 0 \quad\mathrm{if}\quad\zeta_n\not=\zeta_{m},
\\
v^y_{nm} (\kappa_x) = 0 \quad\mathrm{if}\quad\zeta_n=\zeta_{m}.
\end{eqnarray}
\end{subequations}
We close this section by mentioning that the solutions corresponding
to the Dirac point $\bm{K}'$ are analogous to those presented here for
$\bm{K}$. As shown by Marconcini \textit{et al.}~\cite{marconciniKDP},
the wavefunctions for the A sites, Eqs.~\eqref{eq:4A} and
\eqref{eq:5A}, at the $\bm{K}'$ differ by a sign factor from those at
$\bm{K}$. Moreover, the velocity operator at the $\bm{K}'$ has
  the form $\bm{v} = v_F (\sigma_x, -\sigma_y)$. This, together
  with the properties of the envelope functions at both valleys,
  causes the $x$ component of the VME at $\K'$ to have opposite sign
  of those at $\K$; the $y$ components of the VME are the same near
  $\K$ as near $\K'$.
\section{Coherent injection and control\label{sec:CC}}
\subsection{Framework}
In this section, we describe the general framework of the two-color
coherent control scheme. As mentioned in the Introduction, the quantum
interference is between pathways associated with photon absorption
processes arising from different phase related beams. These pathways
connect the same initial and final states, as shown for the processes
in Fig.~\ref{fig:bands_K}, where we consider the two-color scheme with
beams at $\omega$ and at $2\omega$. This figure depicts the two
classes of processes we study in this paper.

The first, \textit{conventional processes}, are those where current
injection arises due to the interference of one-photon
absorption (OPA) at $2\hbar\omega$ and two-photon absorption (TPA) of
(two) photons with energy $\hbar\omega$ \cite{ccontrolDrielSipe2001};
this is depicted with the set of arrows on the right of
Fig.~\ref{fig:bands_K}, under the label ``CONV''. In the remaining of
the discussion, we label variables associated with conventional
processes with a subindex `C'.

The second class of processes arise in experiments on narrow band gap
or gapless materials, with $\hbar\omega>E_g$, where $E_g$ is the
energy band gap.
Under this condition, current injection can arise due to the
interference of OPA at $\hbar\omega$ and stimulated electronic Raman
scattering (ERS) at $\hbar\omega$ \cite{RiouxERS}.
This \textit{ERS} is indicated by the set of arrows at $2\hbar\omega$
and $\hbar\omega$ in the left of Fig.~\ref{fig:bands_K}, under the
label ``ERS''. We refer to variables associated with this Raman
processes with a subindex `R'.  We mention that in coherent control
experiments on typical semiconductors, the beam frequencies employed
are such that $\hbar\omega < E_g< 2\hbar\omega$, and, consequently,
the ERS current is absent because OPA at $\hbar\omega$ is impossible.

Following van Driel and Sipe~\cite{ccontrolDrielSipe2001,
  ccontrolDrielSipe2005}, we calculate the one- and two-photon carrier
injection and current injection rates due to the interaction with a
classical electromagnetic field
\begin{align}
  \bm{E}(t) =   \bm{E}(\omega) e^{-i\omega t} + %
 \bm{E}(2\omega) e^{-2i\omega t} + \mathrm{c.c.},
\end{align}
in the long wavelength limit, where $\omega$ is the fundamental
frequency. 
The interaction between the electric field and the electron system is
accounted by the minimal coupling prescription in the Hamiltonian of
Eq.~\eqref{eq:hamilton}; we do the usual replacement
$p_j\to p_j - e A_j(t)$, for $j=(x,y)$, with
\mbox{$p_j=-i\hbar\partial_j$,} and obtain the interaction Hamiltonian
that acts as the perturbation,
\begin{eqnarray}
  \label{eq:4}
 H_{\mathrm{int}}(t) = - e\, \bm{v}\cdot  \bm{A}(t),
\end{eqnarray}
where $e=-|e|$ is the electron charge and $\bm{A}(t)$ is the vector potential
associated with the electric field, 
$\bm{E}(t)=-\partial \bm{A}(t)/\partial t$.
We treat this problem using standard time-dependent perturbation
theory and Fermi's golden rule. %
Since we are interested in OPA, TPA and ERS processes, the unitary
evolution operator $U(t)$ is expanded perturbatively up to second
order,
\begin{align}\label{eq:evolutionop}
  U(t) &= e^{-i H_0t /\hbar} U_{\mathrm{int}} (t)
\end{align}
where
\begin{align}
   U_{\mathrm{int}} (t) =& 1 + (i\hbar)^{-1} %
\int_{-\infty}^t V_{\mathrm{int}}(t_1) dt_1 \notag\\
&+(i\hbar)^{-2}
\int_{-\infty}^t V_{\mathrm{int}}(t_1) dt_1 %
\int_{-\infty}^{t_1} V_{\mathrm{int}}(t_2) dt_2 +\ldots 
\end{align}
and
\begin{align}
  V_{\mathrm{int}}(t) &= 
e^{i H_0t /\hbar}\,
  H_{\mathrm{int}}(t)\,
e^{-i H_0t /\hbar}.
\end{align} 
Under the perturbation of
Eq.~\eqref{eq:4}, the evolution of the system's state $\big|\Upsilon\rangle$ is not
just the ground state $\big|0\rangle$, but it also contains an
amplitude of the excited state $|nm\bm{\kappa_x}\rangle$ (this ket
corresponds to a state with an electron-hole pair),
\begin{eqnarray}
  \label{eq:1}
  \big|\Upsilon(t)\rangle = c_0(t) |0\rangle + c_{nm\bm{\kappa}_x} (t) |nm\bm{\kappa_x}\rangle
  + \ldots, 
\end{eqnarray}
where $\big|c_{nm\bm{\kappa}_x}(t) \big|^2$ is the probability that
the system is at $\big|nm\bm{\kappa_x}\rangle$; the missing terms in
Eq.~\eqref{eq:1} correspond to higher order excitations, which we
neglect in this work. 
The carrier injection and the current
injection rates are given by
\begin{eqnarray}
  \label{eq:5}
\dot{n} &=&\frac{1}{L} \sum_{nm\bm{\kappa_x}} %
\frac{d}{dt}  \big|c_{nm\bm{\kappa}_x}(t) \big|^2,\\
\dot{J}^a &=&\frac{1}{L} \sum_{nm\bm{\kappa_x}} %
e \left[ v^a_{nn}(\bm{\kappa_x}) - v^a_{mm}(\bm{\kappa_x})\right]\notag\\
&&\times \frac{d}{dt}  \big|c_{nm\bm{\kappa}_x}(t) \big|^2,
  \label{eq:5a}
\end{eqnarray}
respectively, where $L$ is the normalization length introduced below
Eq.~\eqref{eq:sign_edge}. 
To describe the optical processes we are
interested, we compute $c_{nm\bm{\kappa}_x}(t)$ up to
second order (a tutorial derivation can be found in
Ref.~\cite{ccontrolDrielSipe2001}; see also
Ref.~\cite{RiouxERS}). 
Then, the macroscopic expressions for these injection rates get
the form,
\begin{eqnarray}
  \label{eq:5a}
\dot{n}^{(1)} &=& 
\xi^{ab}(\omega)
E^a(-\omega) 
E^b(\omega),
\\
  \label{eq:5b}
\dot{n}^{(2)}_{\rmC} &=&
\xi^{abcd}_{\rmC} (\omega)
E^a(-\omega)
E^b(-\omega)
E^c(\omega)
E^d(\omega),
\\
  \label{eq:5d}
\dot{n}^{(2)}_{\rmR} &=& %
\xi^{abcd}_{\rmR}(\omega) E^{\mathrm{a}}(-2\omega)
 E^{b}(-\omega)
 E^{c}(2\omega)
 E^{d}(\omega),\\
  \label{eq:5c}
\dot{J}^{a}&= &
{\eta}^{\mathrm{abcd}}
(\omega)
\,
E^{b}(-\omega) 
E^{c}(-\omega) 
E^{d}(2\omega) 
+ \mathrm{c.c.},
\end{eqnarray}
where repeated indexes indicate summation, $\omega$ is the
  fundamental frequency, $\dot{n}^{(1)}$ and
$\dot{n}_{\rmC(\rmR)}^{(2)}$ account for the first- and second-order
absorption processes, respectively; overall $\dot{n}$ refers to the
rate of injected carriers per unit length along the ribbon (carriers
per unit length per unit time).  The OPA coefficient is described by a
second-order tensor, $\xi^{ab}$, while the TPA and the ERS absorption
coefficients are described by fourth-order tensors,
$\xi^{abcd}_{\rmC}$ and $\xi^{abcd}_{\rmR}$, respectively.
Here, $\dot{J}^{a}$ includes the electron and hole contributions to
the current (charge per unit time), injected per unit time along the
ribbon. The current injection coefficient $\eta(\omega)$ in
Eq.~\eqref{eq:5c} includes the conventional and the ERS contributions,
i.e., $\eta(\omega)=\eta_C(\omega)+\eta_R(\omega)$.  In the following
sections, we give the full expressions for these coefficients.
Note that the coefficients can be chosen such that 
$\xi^{abcd}_C=\xi^{bacd}_C=\xi^{badc}_C$ and 
$\eta^{abcd}=\eta^{acbd}$.
\subsection{First-order absorption process}
We calculate the expressions for the
coefficients $\xi$ and $\eta$ appearing in
Eq.~\eqref{eq:5a}--\eqref{eq:5c} using  Fermi's golden rule.
For the one-photon absorption coefficient, we obtain
\begin{eqnarray}
  \label{eq:OPA}
      \xi^{\mathrm{ab}}(\omega) &=&
\frac{4\pi e^2}{\hbar^2}
\sum_{nm} \int f_{mn}(\kappa_x)\frac{d{\kappa}_x}{2\pi}
\; \; %
\frac{v_{nm}^{\mathrm{a}} (\kappax)\; v_{nm}^{\mathrm{b}*} (\kappax) %
} {\omega_{nm}^2(\kappax) }%
\nonumber\\
&&\times \delta(\omega_{nm}(\kappax) - \omega), 
\end{eqnarray} 
where we have gone from a sum over states to an integral over
reciprocal space by
$L^{-1}\sum_{\kappa_x}\to(2\pi)^{-1}\int {d\kappa_x}$. 
In this expression %
the sum $\sum_{nm}$ runs over all bands, filled and empty (similarly
for the other response functions considered here);
$\omega_{nm}(\kappax) = \hbar^{-1} E_{nm}(\kappax)$ and
$ E_{nm}(\kappax)= E_n(\kappax)- E_m(\kappax)$ is the energy
difference between two states at a given $\kappax$.
A factor of two has been included to account for spin degeneracy,
which we do throughout the paper. %
The $x-$components of the VME at the $\bm{K}$ and $\bm{K}'$
  valleys differ just by a sign while the $y-$components of the VME
  are the same. Consequently, since all integrals over
  reciprocal space include pairs of VME, the integration over
  $\kappa_x$ can be restricted to a single valley, $\bm{K}$, and
  another factor of two included to account for the contribution of
  the $\bm{K}'$ valley.

\begin{figure}[ht]
  \centering
\includegraphics[scale=.85]{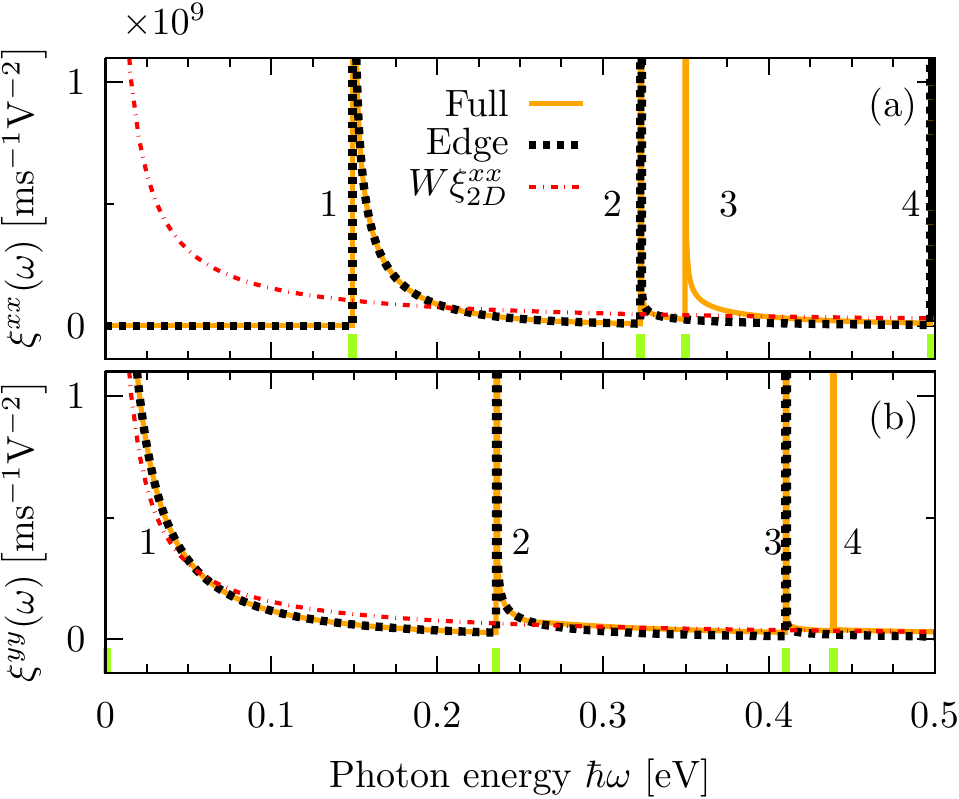}
\caption{(Color online) \label{fig:OPA} One photon absorption spectrum for a ZGNR of
  95 zigzag lines (about 20~nm width). The definitions of the full and
  edge contributions are given in the last paragraph of Sec. II~B. For comparison, we include
  $\Weff\xi^{xx}_{2D}$ (red dot-dashed curves), where $\xi^{xx}_{2D}=\xi^{yy}_{2D}$ is the OPA
  coefficient for graphene~\cite{Rioux2011}, given by
  Eq.~\eqref{eq:opa2Dxx}.
  The green ticks along the horizontal axis indicate photon energies
  at which JDOS divergences occur, which are numbered in concordance
  with Table~\ref{tab:OPApeaks}.}
\end{figure}

The occupation of the states is described by the Fermi-Dirac
distribution. In all of our integrals over reciprocal space
$f_{mn}(\kappax)=f_{m}(\kappax)-f_n(\kappax)$, with
$f_n(\kappa_x) = [1+e^{(E_n(\kappa_x)-\mu)/(k_B T)}]^{-1}$ at
temperature $T$ and chemical potential $\mu$.
Until the end of Sec. IV, we confine ourselves to zero temperature,
hence $f_n(\kappa_x)=\theta(E_n(\kappa_x)-\mu)$, where $\theta$ is the
Heaviside  step function. 
Because of the selection rules for the VME,
Eq.~\eqref{eq:vmeSelectionRules}, the only nonzero components of the
one-photon coefficient are $\xi^{xx}$ and $\xi^{yy}$, which we plot in
Fig.~\ref{fig:OPA} for a system at zero chemical potential. 
As a comparison~\cite{dropoff}, we include
 plots of $\Weff\xi^{xx}_{2D}$, where $W$ is the effective width of
the ribbon,
\begin{align}
\label{eq:opa2Dxx}
  \xi_\mathrm{2D}^{xx}(\omega) = 2\sigma_0  (\hbar\omega)^{-1},
\end{align}
and $\xi_{\mathrm{2D}}^{xx}=\xi_{\mathrm{2D}}^{yy}$ is the OPA
coefficient for a 2D monolayer of graphene\cite{Rioux2011}; here
$\sigma_0=g_s g_v e^2/(16\hbar)$ is the universal optical conductivity
of graphene, and $g_s=2$, $g_v=2$ are the spin and valley
degeneracies, respectively.
For ZGNR, the main difference between the two OPA coefficients is that
$\xi^{yy}$ diverges at zero photon energy, due to a divergence in the
joint density of states (JDOS) between bands $n=+1$ and $n=-1$. In
contrast, for such a pair of bands $\xi^{xx}$ is identically zero, due
to the VME selection rules.  For an undoped ZGNR, $\xi^{xx}$ displays
its first divergence at about 0.15~eV, which is the value of the
band gap at zero Fermi level, and corresponds to the onset of the
transitions $(2,-1)$ and $(1,-2)$ at $\kappax=\Weff^{-1}$; these four
states give the initiation energy for $\xi^{xx}$. In the following we
indicate  a transition from band $m$ to band $n$ by $(n,m)$; hence,
for zero chemical potential and zero temperature, the possible
transitions have $m\leq-1$ and $n\geq 1$. In general, the $\xi^{xx}$
and $\xi^{yy}$ OPA coefficients possess an infinite number of
divergences that arise due to the infinite number of parabolic bands
in the bandstructure. Indeed, the JDOS between states
with band index $n$ and $m$,
\begin{eqnarray}
  \label{eq:jdos}
  \mathrm{JDOS}_{nm} (E) =g_s g_v \int d\kappax \;\delta(E-E_{nm}(\kappax)),
\end{eqnarray}
can be shown to diverge as $(E-E_{nm}^{\mathrm{gap}})^{-1/2}$ for the
confined states, and as $E^{-1}$ for edge states, where
$E$ is the photon energy and $E_{nm}^{\mathrm{gap}}$ is the energy
band gap between bands $n$ and $m$.  In frequency space, these
divergences occur at photon energies $E$ such that
$E=E_{nm}^{\mathrm{gap}}$; in reciprocal space, they occur at
$\kappa_x$ points where argument of the delta function has a zero
derivative. 
The absorption coefficients inherit these JDOS divergences if the
associated velocity matrix elements are nonzero at the $\kappax$ at
which $dE_{nm}/d\kappa_x=0$. The sensitivity of an experiment to these
divergences would depend on the resolution of the photon energy and on
the magnitude of the velocity matrix elements, as well as on the
presence of scattering effects that are not included in this simple
treatment. In every pertaining Figure, we signal the location of these
JDOS divergences by small green ticks.
An interesting characteristic of $\xi^{xx} $ and $\xi^{yy}$ is that
the divergence at the initiation energy always involves an edge state
(see Table~\ref{tab:OPApeaks}); this is reasonable, as these states
are involved in the minimum band gap for an undoped system.

As mentioned above, the sum over states runs over all bands, filled
and empty, but for a given photon energy range (e.g. $0-0.5~e$V, as in
Fig.~\ref{fig:OPA}) the sum requires a finite number of bands. We
refer to this as the ``full'' response.  In order to highlight the
contribution of the edge states, we also compute the response
coefficients with a restricted sum over states $\sum_{nm}$, such that
$n$ or $m$ are $\pm1$, e.g.,
$(n,m)=\{(1,-1), (1,-2), (2,-1), \ldots\}$; we refer to this as the
``edge'' contribution and in the appropriate figures we plot it with
black-dashed lines.
This allows us to easily identify the contribution to OPA from states
at bands $\pm1$.  At low photon energies such contribution is
dominant: for $\xi^{xx}$, all transitions at photon energies
$\hbar\omega<0.350~$eV are from or to edge states; for $\xi^{yy}$, all
transitions at photon energies $\hbar\omega<0.439~$eV are from or to
edge states.  Consequently, at low-photon energies the ``full'' and
``edge'' contributions are indistinguishable. This is shown in
Fig.~\ref{fig:OPA} (see also Table~\ref{tab:OPApeaks}), where for
comparison we also plot $\Weff\xi_{2D}^{xx}$, where $\xi_{2D}^{xx}$ is
the OPA coefficient of graphene calculated~\cite{Rioux2011} at the
same level of approximation adopted here; it is clear how the presence
of the edge states in ZGNR significantly modifies the OPA. Finally, we
mention that the Dirac delta functions appearing in all our
expressions are treated with an interpolation
scheme~\cite{interpNote}.
\newcommand\T{\rule{0pt}{2.6ex}}       
\newcommand\B{\rule[-1.2ex]{0pt}{0pt}} 
\begin{table}[ht]
  \centering
  \caption{\small Onset energies for the lowest energy transitions for
    an undoped cold ZGNR. Tuples $(n,m)$ indicate a transition
    from band $m$ to band $n$  and every onset energy indicates the position 
    of a JDOS divergence. The peak number is as indicated in
    Fig.~\ref{fig:OPA}.}
  \label{tab:OPApeaks}
  \begin{tabular}{ c  c  c  c  c }
    \hline
    \hline
Peak& \multicolumn{2}{c}{$\xi^{xx} $} & \multicolumn{2}{c}{$\xi^{yy} $}  \T\B\\
    \cline{2-5}
number\hspace*{2mm}  & E (eV) & Transition &  E (eV) & Transition \T\B\\
    \hline
    1 & 0.149 & $(2,-1), (1,-2)$ & 0.000 &$(1,-1)$ \\
    2 & 0.323 & $(4,-1), (1,-4)$ & 0.236 &$(3,-1), (1,-3)$\\
    3 & 0.350 & $(3,-2), (2,-3)$ & 0.410 & $(5,-1), (1,-5)$\\
    4 & 0.498 & $(6,-1), (1,-6) $ & 0.439& $(4,-2), (2,-4) $\\
    \vdots &     \vdots &     \vdots &  \vdots& \vdots\\
    \hline
  \end{tabular}
%
%
\end{table}

\subsection{Second-order absorption processes}
\subsubsection{Conventional process}
\begin{figure}[ht]
  \centering
\newcommand{\LAESCALA}{2.5}
  \includegraphics[scale=\LAESCALA]{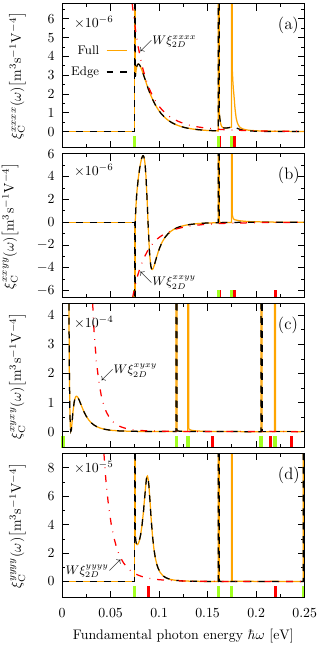}
  \caption{(Color online) Nonzero two photon absorption coefficients $\xi^{abcd}_C$
    for a ZGNR with 95~zigzag-lines (about 20~nm width). The definitions of the full and
  edge contributions are given in the last paragraph of Sec.~II~B. On each
    panel, we include $\Weff\xi^{abcd}_{2D}$ (red dot-dashed curves), where
    $\xi^{abcd}_{\mathrm{2D}}$ (Eq.~\eqref{eq:tpa2Dxxxx} and text below) is for a
    graphene sheet \cite{Rioux2011}.  The green (red) ticks along the
    horizontal axis indicate the photon energies at which JDOS
    divergences (resonances) occur.}
  \label{fig:TPA}
\end{figure}

In this section, we start by considering the second order process
related to the absorption of two photons of energy $\hbar\omega$,
indicated by the rightmost arrows in Fig.~\ref{fig:bands_K}.  Carrying
the perturbation calculation up the second order, we obtain the
two-photon absorption (TPA) coefficient,
\begin{eqnarray}
  \label{eq:8}
  \xi^{\mathrm{abcd}}_C (\omega)
  &=& 
\frac{64\pi e^4}{\hbar^4} 
\sum_{nm}  \int f_{mn} \frac{d\kappa_x}{2\pi} \;
\frac{
\mathcal{V}_{C;nm}^{a b*}
\;
{\mathcal{V}}_{C;nm}^{ c d}}  %
{ \omega_{nm}^4(\kappax)}
\nonumber\\
&&\times
\delta(\omega_{nm}(\kappax) -2\omega),
\label{eq:TPAfinal}
\end{eqnarray}
where 
\begin{eqnarray}
  \label{eq:effVME}
    {\mathcal{V}}_{C;nm}^{\, \rmi\,  \rmj} \equiv %
\hbar \; 
\sum_{\ell}
\frac{  %
v_{n\ell }^{\rmi}  \;  v_{\ell m}^{\rmj}  %
+ %
v_{n\ell }^{\rmj} \; v_{\ell m}^{\rmi}
} %
{ 2E_{\ell} - E_{n} -E_{m} + i \beta_C },
\end{eqnarray}
which we regard as the \textit{effective velocity matrix element}
(effective VME) for the second order conventional process (C)
process. Here $\beta_C$ is a small constant introduced to
  broaden resonant processes (discussed below) and the sum over
  $\ell$ corresponds to the virtual electron and virtual hole
  contributions~\cite{ccontrolDrielSipe2001}. Although this sum runs
  over all bands (filled and empty), a converged value is obtained for
  $\ell=20$ bands for a photon energy range of \mbox{0--1~$e$V}. From
the selection rules for the regular VME,
Eq.~(\eqref{eq:vmeSelectionRules}), we obtain the selection rules for
$\mathcal{V}$,
\begin{subequations}
\label{eq:vmeSelectionRulesEFF}
\begin{align}
\mathcal{V}^{xx}_{C;nm}
  &=0\quad\mathrm{if}\quad\zeta_n\not=\zeta_{m},\\
\mathcal{V}^{yy}_{C;nm}
 &=0\quad\mathrm{if}\quad\zeta_n\not=\zeta_{m}, \\
\mathcal{V}^{xy}_{C;nm}
  &=0\quad\mathrm{if}\quad\zeta_n =\zeta_{m},
\end{align}
\end{subequations}
and from this we identify eight nonzero $\xi^{abcd}_C$
  components, four of them independent, namely
$ \xi^{xxxx}_C,
    \xi^{xxyy}_C=\left(\xi^{yyxx}_C\right)^*,
    \xi^{xyxy}_C=\xi^{xyyx}_C=\xi^{yxxy}_C=\xi^{yxyx}_C,$
and $\xi^{yyyy}_C$, which we show in Fig.~\ref{fig:TPA}.  
A  feature of these coefficients is that the onset of the
two-photon absorption signal is at the minimum band gap between bands
$(2,-1)$, except for $\xi^{xyxy}_C$, which has its onset at
0~\textit{e}V; this follows from the selection rules for the effective
VME, which are inherited from the usual VME, and indicate that the
transition $(1,-1)$ is allowed.

As we found for the OPA coefficients $\xi^{ab}$, the TPA coefficients
$\xi^{abcd}_C$ suffer from divergences, but for the TPA coefficients
they are of two types: JDOS divergences and
effective-VME-divergences. The latter results when the nominal virtual
state lies at the average of the energies between two transition
states, $|n\kappax\rangle$ and $|m\kappax\rangle$, i.e., when (see
Eq.~\eqref{eq:effVME})
\begin{eqnarray}
  \label{eq:divergency}
  E_{\ell} = ( E_{n} + E_{m}) /2.
\end{eqnarray}
Such condition corresponds to a resonant TPA and an instance
where this occurs is indicated on Fig.~\ref{fig:bands_K} by the three
dots along the vertical line at $\kappa_x=0$. In Fig.~\ref{fig:TPA} we
distinguish these two types of divergences by small vertical lines of
different color; a green tick indicates the presence of a
JDOS-divergence, while a red tick indicates the presence of an
effective-VME-divergence. 
In order to broaden the latter resonances, a small damping constant
$\beta_C$ of 20~m\textit{e}V was introduced in the denominator of
Eq.~\eqref{eq:effVME}.  This value, which is close to the thermal energy
$k_BT$ associated with room temperature, was chosen arbitrarily. A
more detailed theory would be necessary to indicate how these
resonances are really broadened; the choice we make here simply allows
us to identify easily where these resonances occur in our
calculations.
We mention that the onset of $\xi^{xxxx}_C$ is due to the transitions
$(2,-1)$ and $(1,-2)$, which are free from resonances because the
matrix elements to the intermediate states (one of the edge bands
$\pm1$ that would lead to a divergent condition) are forbidden
by the selection rules. 
Therefore, in the photon energy range 0 to 0.15~eV, the
coefficient $\xi^{xxxx}_C$ is free of resonances.

We present the $\xi^{abcd}_C$ coefficients in Fig.~\ref{fig:TPA}, and
identify the edge contributions to them (black-dashed lines). As we
found for $\xi^{ab}$, for $\xi^{abcd}_C$ the edge states make a
dominant contribution at low photon energies, and are involved at the
onset of TPA. As a comparison~\cite{dropoff}, in Fig.~\ref{fig:TPA}, we include
 plots of $\Weff\xi^{abcd}_{2D}$, where $W$ is the effective width of
the ribbon,
\begin{align}
\label{eq:tpa2Dxxxx}
\xi_{\mathrm{2D}}^{xxxx} (\omega)= 8 g_s g_v \hbar e^4 v_F^2
                                     (2\hbar\omega)^{-5},
\end{align}
and
$\xi_{\mathrm{2D}}^{xxxx}=\xi_{\mathrm{2D}}^{yyyy}=\xi_{\mathrm{2D}}^{xyxy}=\xi_{\mathrm{2D}}^{xyyx}=-\xi_{\mathrm{2D}}^{xxyy}$
are the TPA coefficients for a 2D monolayer of
graphene~\cite{Rioux2011}; as before,  $g_s=2$ and  $g_v=2$ are the spin and
valley degeneracies, respectively. 

\subsubsection{ERS process}
\begin{figure*}[ht]
  \centering
\newcommand{\LAESCALA}{2.5}
  \includegraphics[scale=\LAESCALA]{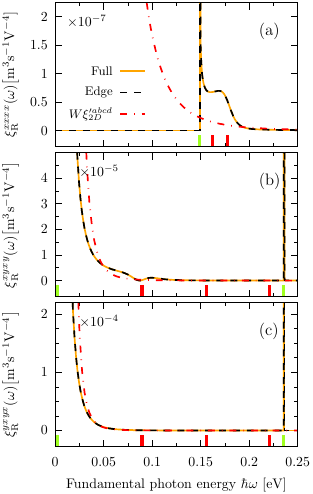}\hspace*{10mm}
  \includegraphics[scale=\LAESCALA]{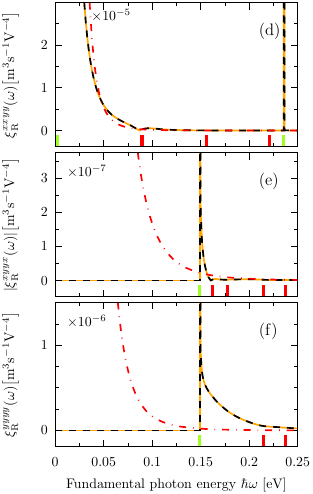}
  \caption{(Color online) ERS carrier injection tensor, as given by
    Eq.~\eqref{eq:TPAfinalERS}. The definitions of the full and edge
    contributions are given in the last paragraph of Section
    II.B. Notice that the edge states play a dominant contribution to
    the ERS absorption process, due to the large amount of resonant
    states. The green (red) ticks along the horizontal axis indicate
    the photon energies at which JDOS divergences (resonances) occur.
    The red dot-dashed lines indicate the ERS processes for 2D
    graphene~\cite{RiouxERS}.}
  \label{fig:ERStpa}
\end{figure*}

Now we consider another second order process involving light at
$2\omega$ and light at $\omega$, stimulated electronic Raman
scattering, which can be characterized as virtual absorption at
$2\hbar\omega$ followed by emission at $\hbar\omega$; see the left
diagram in Fig.~\ref{fig:bands_K}. %
This process exists in semiconductors when the fundamental photon
energy is larger than the band gap, which is always the case for an
undoped ZGNR, because the edge states provide a zero-gap system. %
Following an earlier treatment of graphene \cite{RiouxERS}, we find
the ERS carrier injection to be
\begin{eqnarray}
  \label{eq:80}
  \xi^{\mathrm{abcd}}_{\rmR} (\omega)
  &=& 
\frac{2\pi e^4}{\hbar^4} 
\sum_{nm}  \int f_{mn} \frac{d\kappa_x}{2\pi}
\frac
{
{\mathcal{V}}_{\rmR;nm}^{\, d\, a\,*} %
{\mathcal{V}}_{\rmR;nm}^{\, b\, c}
} 
{\omega_{nm}^4}
\notag\\
&&\times\delta(\omega_{nm}(\kappax) -\omega), \quad
\label{eq:TPAfinalERS}
\end{eqnarray}
where the \textit{effective VME} for the  ERS process are
\begin{eqnarray}
  \label{eq:effVMEers}
    {\mathcal{V}}_{\rmR; nm}^{\, \rmi\,  \rmj} \equiv %
\hbar 
\sum_{\ell}
\Big[
\frac{v_{n\ell }^{\rmi} v_{\ell m}^{\rmj} } {E_{\ell n}-E_{nm}+i\beta_\rmR}
+\notag\\
+\frac{  v_{n\ell }^{\rmj} \; v_{\ell m}^{\rmi} } {E_{\ell m} + E_{nm}+i\beta_\rmR}
\Big].
\end{eqnarray}
As in Eq.~\eqref{eq:effVME}, $\beta_\rmR$ is a small constant
  introduced to broaden resonant processes and the sum over $\ell$
  runs over all bands (filled and empty), but a converged value is
  obtained for $\ell=30$ bands for a photon energy range of
  \mbox{0--1~$e$V}. The first term in the sum of
Eq.~\eqref{eq:effVMEers} corresponds to photo-emission by an electron,
and the second to photo-emission by a hole \cite{RiouxERS}.
Note that due to the different frequencies involved in
  Eq.~\eqref{eq:5d}, symmetrization of
  $\mathcal{V}_{\rmR}^{\mathrm{ij}}$ is unnecessary.
The selection rules for $\mathcal{V}_{\rmR}^{\mathrm{ij}}$ are the
same as those for $\mathcal{V}_{\rmC}^{\mathrm{ij}}$
(Eq.~\eqref{eq:vmeSelectionRulesEFF}); note, however, that 
$\mathcal{V}_{\rmR}^{\mathrm{ij}}\not=\mathcal{V}_{\rmR}^{\mathrm{ji}}$,
although $\mathcal{V}_{\rmR}^{\mathrm{ij}}$ and
$\mathcal{V}_{\rmR}^{\mathrm{ji}}$ satisfy the same selection
rule. 
From this we identify six nonzero
terms for the ERS carrier
injection coefficient,
$\xi _{\rmR}^{xxxx}$,  $\xi _{\rmR}^{xyyx}=\left(\xi _{\rmR}^{yxxy}\right)^*$,
$\xi _{\rmR}^{xxyy}=\left(\xi _{\rmR}^{yyxx}\right)^*$, $\xi _{\rmR}^{xyxy}$,
$\xi _{\rmR}^{yxyx}$, and $\xi _{\rmR}^{yyyy}$.
As do the conventional coefficients, the ERS coefficients 
suffer from JDOS and effective-VME divergences, the later arising
whenever 
\begin{subequations}
\begin{eqnarray}
  \label{eq:ers_divs}
\label{eq:div_electron}
  E_\ell &=& 2 E_n - E_m \quad \text{or} \\
\label{eq:div_hole}
E_\ell &=& 2E_m - E_n
\end{eqnarray}
\end{subequations}
is satisfied. These conditions correspond to resonant processes,
when a state is located at  an energy $|E_{nm}(\kappa_x)|$ above (below) the final
(initial) state $n$ ($m$). As in  Eq.~\eqref{eq:effVME}, a
small damping constant $\beta_R$ of 20~m\textit{e}V was
introduced in the denominators of Eq.~\eqref{eq:effVMEers}.
All of these ERS coefficients present a large number of these
resonances, causing $\xi^{abcd}_{\rmR}$ to be highly sensitive to the
value of the $\beta_{\rmR}$ parameter. However, these resonances are
of small magnitude for the energy range chosen for
Fig.~\ref{fig:ERStpa}, hence they are not apparent. 
As shown, three of these components have their onset at zero photon energy,
because the symmetry properties of the involved matrix elements allow
for transitions between the two edge states.

\subsection{Current Injection}
\subsubsection{Injection coefficients}
We begin with the expression for $\eta_C$, the current injection
coefficient characterizing the conventional process. Here the
interference between the TPA at $\hbar\omega$ with OPA at
$2\hbar\omega$ (see the right diagram in Fig.~\ref{fig:bands_K}) leads
to net current injection coefficients (including electron and hole
contributions) given by~\cite{ccontrolDrielSipe2001}
\begin{eqnarray}
  \label{eq:13}
  \eta_{\rmC}^{\mathrm{abcd}} (\omega) 
&=&
\frac{16 i \pi e^4 } {\hbar^3}%
\sum_{nm}\int f_{m n}
\frac{d \kappa_x} {2\pi} 
\frac{
\left( v_{nn}^{\mathrm{a}} 
-
v_{mm}^{\mathrm{a}} 
\right)
    \mathcal{V}^{bc*}_{\rmC; nm}
v_{nm}^{\mathrm{d}} 
}
{\omega_{nm}^3}
\nonumber\\
&&
\times\delta( \omega_{nm}(\kappax)-2\omega).
\end{eqnarray}
From the selection rules for the regular and the effective VME,
Eq.~\eqref{eq:vmeSelectionRules} and
Eq.~\eqref{eq:vmeSelectionRulesEFF}, we identify %
three nonzero current injection coefficients,
$\eta_\rmC^{xxxx}, \eta_\rmC^{xyyx},$ and
$\eta_\rmC^{xxyy}=\eta_\rmC^{xyxy}$.  Notice that for all these
tensors the first Cartesian component is $x$: Due to the confinement of the
  ribbons along the $\hat{y}$ direction (see
  Fig.~\ref{fig:structure}), the current injection can only flow along
  the $\hat{x}$ direction, and all tensor components
  $\eta_\rmC^{yabc}$ are zero.
\begin{figure}[ht]
  \centering 
  \includegraphics[scale=2.5]{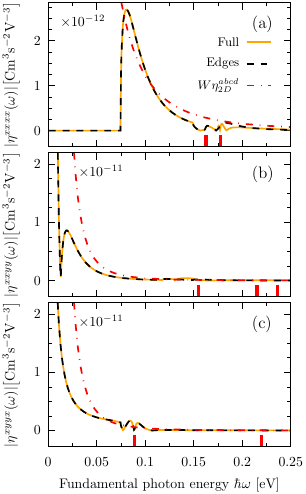}
  \caption{(Color online) Nonzero net current injection coefficients,
    including the conventional and ERS contributions, i.e.
    $\eta(\omega)=\eta_{\rmC}(\omega)+\eta_{\rmR}(\omega)$.  The definitions of the full and
  edge contributions are given in the last paragraph of Section II.B. On each
    panel, we include $\Weff\eta^{abcd}_{2D}$ (dot-dashed red curves),
    where $\eta^{abcd}_{\mathrm{2D}}$ (Eq.~\eqref{eq:eta2Dxxxx} and
    text below) is for a graphene sheet \cite{RiouxERS}.
    The red ticks along the horizontal axis indicate the energies at
    which resonances occur; a damping constant of 20~meV is introduced
    to broaden such resonances. The dips observed in these
    coefficients arise due to negative contributions to the
    conventional and ERS currents, in turn due to the shape of the
    involved matrix elements. }
  \label{fig:CURR} 
\end{figure}
Turning to the expression for $\eta_R$, the current injection
coefficient characterizing the interference between the ERS discussed
above and the OPA at $\omega$ (see the left diagram in
Fig.~\ref{fig:bands_K}), including both electron and hole
contributions we find
\begin{eqnarray}
  \eta_\rmR^{abcd} (\omega)= %
\frac{2 i \pi e^4 }{\hbar^3} %
 \sum_{nm} \; \int f_{m n}\frac{d\kappa_x}{2\pi}  %
 \frac{ \left( v^{a}_{nn} -v^{a}_{mm} \right) }
{\omega^3_{nm}}
\notag\\ 
\qquad\times\Big[v^{b*}_{nm }  \mathcal{V}^{cd}_{R;nm} +
  v^{c*}_{nm }   \mathcal{V}^{bd}_{R;nm} \Big] \;
  \delta(\omega_{nm}(\kappax) - \omega),\quad
\label{eq:20}
\end{eqnarray}
where $\mathcal{V}_\rmR$ is given by Eq.~\eqref{eq:effVMEers}. On the basis of the
matrix elements selection rules, we identify three nonzero ERS current
injection coefficients, $\eta_\rmR^{xxxx}$, $\eta_\rmR^{xyyx}$, and
$\eta_\rmR^{xxyy} = \eta_\rmR^{xyxy}$.

Over the frequency range shown in Fig.~\ref{fig:CURR}, the
conventional and the ERS current injection coefficients are of the
same order, dropping off as the inverse of the third power of the
photon energy, as do the coefficients for
graphene~\cite{RiouxERS}. Thus we only plot the total injection
coefficients $\eta^{abcd}=\eta_\rmC (\omega) +\eta_\rmR (\omega)$. 
For comparison, we include plots of $\Weff\eta^{abcd}_{2D}$ (with the
respective values of the Cartesian indices), where
\begin{align}
\label{eq:eta2Dxxxx}
  \eta_\mathrm{2D}^{xxxx}(\omega) = i \frac{3}{4} g_s g_v e^4 v_F^2 (2\hbar\omega)^{-3},
\end{align}
and
$\eta_{\mathrm{2D}}^{xxxx}=3\eta_{\mathrm{2D}}^{xxyy}=3\eta_{\mathrm{2D}}^{xyyx}$
are the \textit{net} current injection coefficients for a 2D monolayer
of graphene\cite{RiouxERS}; as before, $g_s=2$ and $g_v=2$ are the spin and
valley degeneracies, respectively. 
As we saw for carrier injection, the edge states provide the strongest
contribution at the onset of current injection. Another characteristic
of these coefficients is that $\eta^{xxxx}$ has its onset at the
band gap between bands $(2,-1)$, while ${\eta}^{xxyy}$ and
${\eta}^{xyyx}$ have their onset at 0~eV.  This is due to the
selection rules that the matrix elements involved in both the
conventional and ERS process satisfy, allowing transitions between
bands $(1,-1)$.
An important characteristic of the current injection coefficients is
that they are free of JDOS divergences, because the diagonal matrix
elements in their respective expressions, Eqs.~\eqref{eq:13} and
\eqref{eq:20}, are identically zero at the $\kappax$ at which the
minimum gap occurs.  %
However, a number of effective VME resonances do exist at photon
energies indicated by the small red ticks in Fig.\ref{fig:CURR}, such
that Eq.~\eqref{eq:divergency} is satisfied. As explained before, the
magnitude of these resonances is broadened by a small damping
constant.
These coefficients are shown in Fig.~\ref{fig:CURR}, where we present
the net current injection arising from the addition of the
conventional and ERS contributions,
i.e., $\eta(\omega)=\eta_\rmC(\omega)+\eta_\rmR(\omega)$.
%
%
\subsubsection{Swarm velocities}
\begin{figure}[ht]
  \centering
  \includegraphics[scale=0.8]{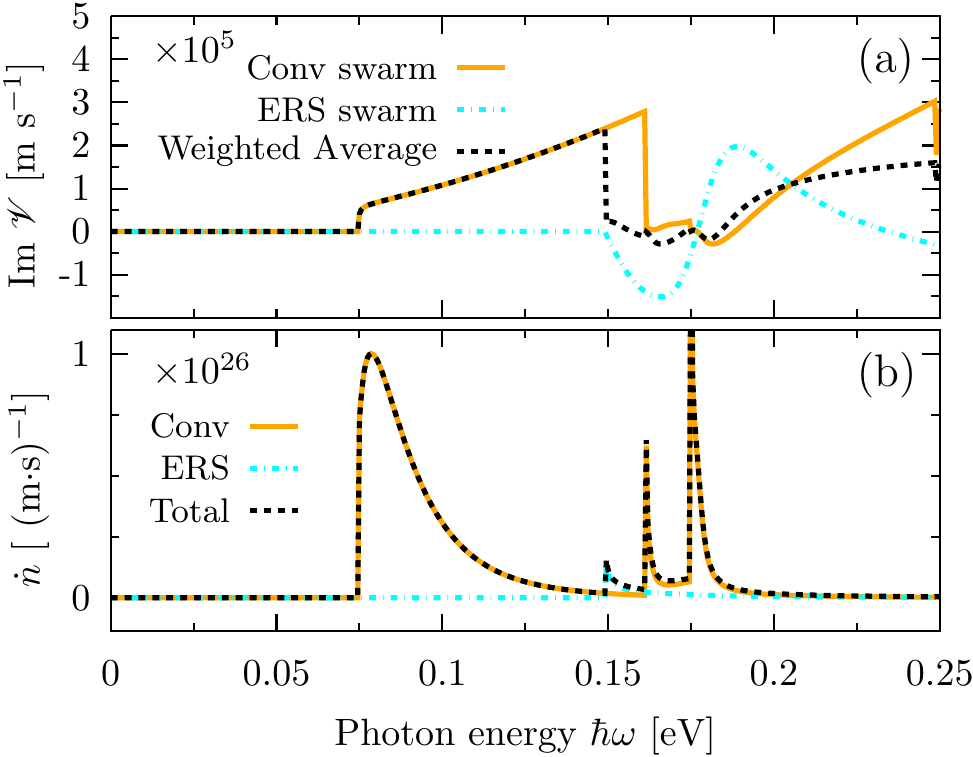}
  \caption{(Color online) Swarm and (weighted) average velocities
    (top), accompanied by the carrier density rates (bottom)
    along $\xhat$ due to $\eta^{xxxx}$ for
    typical~\cite{sun-norris2010} beam intensities of the fundamental
    and second harmonic fields. The average carrier velocities
    (black-dashed line) for $\eta^{xxyy}$ and $\eta^{xyxy}$ are of the
    same order, but their net components have a smooth onset at zero
    photon energy.}
  \label{fig:swarm}
\end{figure}

The numerical values of the coefficients $\xi^{ab}$,
$\xi^{abcd}_{\rmC(\rmR)}$, and $\eta^{abcd}_{\rmC(\rmR)}$ do not
immediately give a sense of the average velocities with which the
electrons and holes are injected. Sometimes an average, or
\textit{swarm} velocity is introduced to indicate
this~\cite{ccontrolDrielSipe2001}. In the system considered here, we
could introduce a swarm velocity for both the conventional and ERS
processes, according to
\begin{align}
\label{eq:swarm}
\bm{\Vswarm}_{\rmC(\rmR)} &= 
  \frac{1}{e}\;\;\frac{\bm{\dot{J}}_{\rmC(\rmR)}(\omega)}{\dot{n}^{(1)}(\Omega) +
    \dot{n}_{\rmC(\rmR)}^{(2)}(\omega) },
\end{align}
where $\Omega=2\omega$ for $\bm{\Vswarm}_{\rmC}$ because
$\bm{\dot{J}}_{\rmC}$ arises from the interference of OPA at $2\omega$
with TPA at $\omega$, while $\Omega=\omega$ for
$\bm{\Vswarm}_{\rmR}$ because $\bm{\dot{J}}_{\rmR}$ arises from the
interference of OPA at $\omega$ with the ERS described above.
Besides describing an average speed that characterizes the injected
carriers, one can consider maximizing Eq.~\eqref{eq:swarm} by using
appropriate phases in the optical beams, and adjusting the relative
amplitudes of the light at $\omega$ and $2\omega$. 
Considering just the swarm velocity of the conventional process, such
optimization leads to equal OPA and TPA, and it follows
that the intensity of the fundamental beam at $\omega$ should be about
half an order of magnitude larger that of the beam at $2\omega$, for a
fundamental photon energy of about $0.4~e$V.
In contrast, the swarm velocity of the ERS process depends only on the
intensity of the beam at $2\omega$. Further, in trying to optimize the
\textit{net} swarm velocity, determined by the total current injected
divided by the total carrier density injected, one finds that the beam
at $2\omega$ should have an intensity about an order of magnitude
larger than the beam at $\omega$. 
Since in typical experiments the beam at $2\omega$ is obtained by
second harmonic generation of part of the beam at $\omega$, this would
be impractical.
Thus we calculate the conventional and Raman swarm velocities for
typical~\cite{sun-norris2010} beam intensities of the fundamental and
second harmonic fields, shown in Fig.~\ref{fig:swarm}. We complement
these carrier velocities with the total average velocity of the
injected carriers
\begin{align}
  \bar{\bm{\Vswarm}}_{\mathrm{tot}} &= 
\frac{1}{e} \; %
\frac{ %
   \bm{\dot{J}}_{\rmR}(\omega)+
  \bm{\dot{J}}_{\rmC}(\omega)
}
{ %
  \dot{n}^{(1)}(\omega) +
  \dot{n}^{(1)}(2\omega) +
  \dot{n}_{\rmC}^{(2)}(\omega) +
  \dot{n}_{\rmR}^{(2)}(\omega)
},
\end{align}
also evaluated at typical \cite{sun-norris2010} beam
intensities. These carrier velocities are shown in
Fig.~\ref{fig:swarm}.  
As a reference, at the photon energy of 0.25~\textit{e}V, the maximum
swarm velocity of the conventional process for a monolayer of graphene is
$2.9\times10^{5}$~ms$^{-1}$. Hence the carrier velocities in ZGNR are
comparable to those on a monolayer of graphene, as might be expected.
%
%
%
%
%
%
%
\section{Doping \label{sec:doping}}
In the previous sections, we investigated the carrier and current
injection at zero chemical potential.  Since the dispersion relations
of the edge states in ZGNR have a zero band gap and are flattened for
$\kappa_x>W^{-1}$ (Fig.~\ref{fig:bands_K}), those states are always
involved at the onset energy of all of the optical response
coefficients studied here. This suggests that doping is an effective
method to alter the population of these two bands and the current that
can be injected by the optical transitions between them. In this
section, we revisit the calculations of $\xi^{ab}$, $\xi^{abcd}_{C(R)}$
and $\eta^{abcd}$ for a negative chemical potential, corresponding to
a $p$-doped system.
Besides the modified contribution from the edge states, we will also
see significant modification in the contributions from other bands,
particularly in the region near the $\mathbf{K}$ and $\mathbf{K}'$
points, where doping leads to either a ``valley'' of filled states
($n$-doped), or a ``hill'' of unfilled states ($p$-doped); see
Fig.~\ref{fig:bands_K}.

We consider two negative Fermi levels, $\mu_1=-0.1$~\textit{e}V and
$\mu_2=-0.2$~\textit{e}V, which in Fig.~\ref{fig:bands_K} we
  indicate by the upper boundaries of the grey areas.  The value of $-0.1$~\textit{e}V
is interesting because, at this chemical potential, the flat part of
band $-1$ (i.e., the region where $\kappax>\Weff^{-1}$,
cf. Fig.~\ref{fig:bands_K}) contains empty states; this condition
allows transitions from lower energy bands with final states in band
$-1$, but also disables transitions from band $-1$ to upper bands.
The second value, $\mu=-0.2$~\textit{e}V, is interesting because at this
potential a ``hill'' of unfilled states arises in the first parabolic
band (band $-2$ in Fig.~\ref{fig:bands_K}) 
at energies below our nominal value of zero.

\begin{figure}[ht]
  \centering
\includegraphics[scale=0.85]{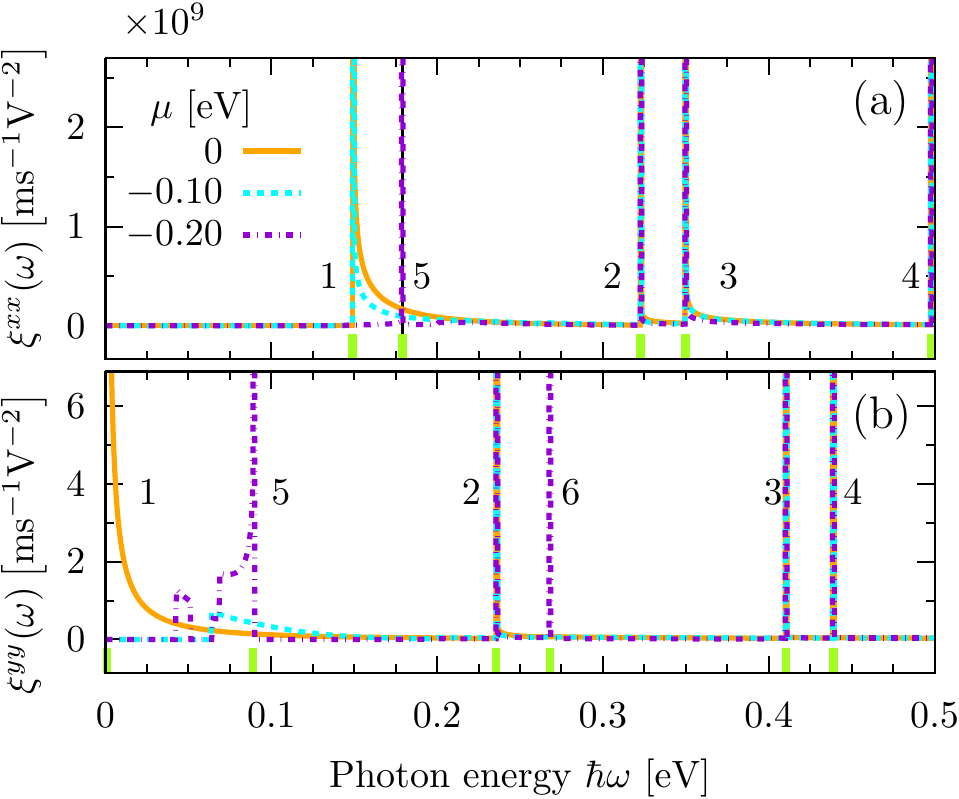}
\caption{(Color online) \label{fig:OPAfermis} One photon absorption
  coefficients as a function of the photon energy for selected Fermi
  levels corresponding to p-doped samples. The ZGNR has 95 zigzag
  lines (about $20$~nm width).  For nonzero chemical potentials, some
  transitions become impossible and some new transitions arise,
  possibly leading to new JDOS divergences (e.g. divergences 5 and
  6). Divergences 1--4 are the same as in Fig.~\ref{fig:OPA}.  }
\end{figure}

We present the results of the calculations of OPA coefficients for
those values of the chemical potential in Fig.~\ref{fig:OPAfermis}.
In an undoped sample, the JDOS divergences in $\xi^{xx}$ at low photon
energies are due to the onset of the transitions
$(2,-1), (1,-2), (4,-1),$ and $(1,-4)$ (see Table~\ref{tab:OPApeaks}
and Fig.~\ref{fig:OPA}). Since all of these transitions involve bands
$\pm1$, any nonzero chemical potential has the capacity to
significantly alter the OPA at these photon energies.
For instance, if the Fermi level is at $-0.1$~eV, then the flat
part of band $-1$ contains empty states, and the low
photon energy divergences are removed. In
addition, at this chemical potential transitions of the type
$(-1, n)$, for $n$ odd and $<-1$ are permitted.
However, the contributions to the OPA from these new
transitions are of smaller magnitude than the contribution from the
$(1,-2)$ transition, which is unaffected by the $-0.1$~eV doping.  For
this reason, the $(1,-2)$ transition remains as the main contribution
to the $\xi^{xx}$ coefficient at low photon energies at this chemical
potential (see
Fig.~\ref{fig:OPAfermis}). \\

At the Fermi level $-0.2$~eV, the edge states are completely
empty, as are the states at the higher points of band $-2$ near the
$\mathbf{K}$ and $\mathbf{K}'$ points.  This condition allows
transitions of the type $(-2, n)$, for $n$ even and $<-2$, and also
forbids transitions of the type $(n, -2)$, for $n$ odd and $\geq 1$,
and $\kappa_x$ near the $\mathbf{K}$ and $\mathbf{K}'$ points. It is
this latter restriction which significantly changes the $\xi^{xx}$
coefficient near its onset. %
A further decrease in the Fermi level would consistently remove the
divergences in $\xi^{xx}$ at low photon energies. All these
observations were confirmed with a band-by-band calculation of
$\xi^{xx}$.

The effect of doping the system has a larger influence on the onset
energy of $\xi^{yy}$ that on that of $\xi^{xx}$. This is because the
JDOS divergences at low photon energies relevant for $\xi^{yy}$ are
due to
the transitions $(1,-1)$, $(3,-1),$ and $(1,-3)$
(cf. Table~\ref{tab:OPApeaks}).
Therefore, even for small doping, the large contribution coming
from the transitions between the two edge states (bands $\pm1$,
$\kappa_x>\Weff^{-1}$) is significantly decreased, 
and leads to a greater change of the magnitude of $\xi^{yy}$ than of
the magnitude of $\xi^{xx}$. %
A special signature of $\xi^{yy}$ %
for $\mu=-0.2~e$V (dark-violet signal, Fig.~\ref{fig:OPAfermis} b)) is
the presence of two narrow peaks at 0.045 and 0.075~\textit{e}V; %
the first of these peaks is due to the $(-1,-2)$ transition, while the
second is from the $(-2,-3)$ transition. These two transitions are active
only for those $\kappa_x$ states at which the ``hill'' of band $-2$ is
empty (see Fig.~\ref{fig:bands_K}).  
Notably, the transition $(-2,-3)$ brings a new JDOS divergence because
it is active over a range of reciprocal space that includes
$\kappa_x=\Weff^{-1}$, where both bands have their maximum and their
energy difference $E_{nm}(\kappa_x)$ has a zero derivative (see the
discussion below Eq.~\eqref{eq:jdos}).

In general, all these new transitions involve more JDOS
  divergences if the range of $\kappax$ over which they are active
includes the $\kappax$ at which the band pairs have their maxima or
minima. For instance, the divergences 1--4 in Fig.~\ref{fig:OPAfermis}
are the same as those in Fig.~\ref{fig:OPA} and
Table~\ref{tab:OPApeaks}, but the divergences 5--6 arise due to the
new transitions allowed at nonzero chemical potentials: in
\mbox{Fig.~\ref{fig:OPAfermis}~a)}, at the chemical potential
$-0.20~e$V, the divergence 5 at $0.179$~eV is due to the transition
$(-2,-4)$, which is active over a range of $\kappa_x$ that includes
the $\kappa_x$ at which bands $(-2,-4)$ have their maxima, hence a new
JDOS divergence appears.  Likewise for $\xi^{yy}$ in
\mbox{Fig.~\ref{fig:OPAfermis}~b)} at $\mu=-0.20~e$V: divergences 5 at
$0.089$~eV and 6 at $0.268$~eV exist because the transitions $(-2,-3)$
and $(-2,-5)$ are active over regions of reciprocal space that include
the $\kappa_x$ at which such bands have their maxima.

\begin{figure}[ht]
  \centering
\newcommand{\LAESCALA}{2.5}
\includegraphics[scale=\LAESCALA]{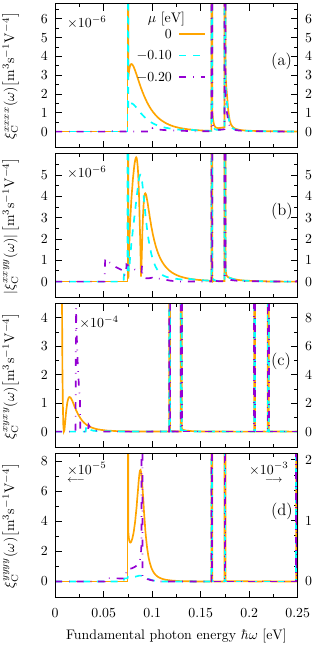}
%
%
\caption{(Color online) \label{fig:TPAfermis} Two photon absorption
  coefficients for selected Fermi levels corresponding to $p$-doped
  samples. The ZGNR has 95 zigzag lines (about $20$~nm width). For
  panels where two different vertical scales are present, i.e. panel 
  (d), the scale on the left (right) is for undoped (doped) cases
  (arrows below the factors indicate the ordinate for which they
  apply). A damping constant $\beta_C=20~$meV was introduced.}
\end{figure}

\begin{figure*}[ht]
  \centering
\newcommand{\LAESCALA}{2.5}
  \includegraphics[scale=\LAESCALA]{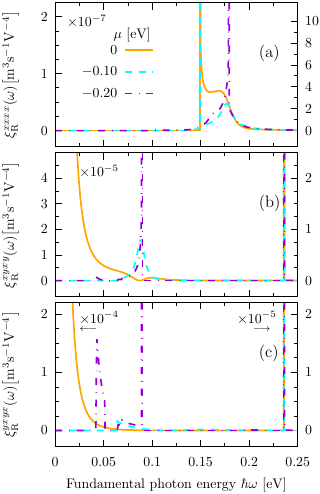}
\hspace*{10mm}
  \includegraphics[scale=\LAESCALA]{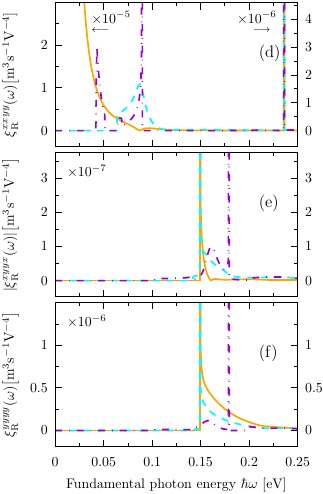}
  \caption{(Color online) ERS carrier injection  coefficients for selected Fermi levels
  corresponding to p-doped samples.  The ZGNR has 95
  zigzag lines (about $20$~nm width). For panels where two different
  vertical scales are present, i.e. (c) and  (d), the scale on the left
  (right) is for undoped (doped) cases (arrows below the factors
  indicate the ordinate for which they apply). Notice that at $\mu=-0.20$~meV some resonances are
    absent, e.g., at 0.15~eV in (a) and (f); this is because, at this
    Fermi level, the states at which these resonances are present for
    the undoped system, now contain empty states. A
  damping constant $\beta_C=20~$meV was introduced.
 }
  \label{fig:tpaers_fermis}
\end{figure*}

\begin{figure}[ht]
\newcommand{\LAESCALA}{2.5}
  \centering
  \includegraphics[scale=\LAESCALA]{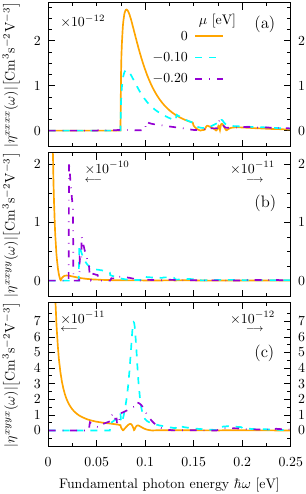}
  \caption{(Color online) Net current injection tensors (conventional
    plus ERS contributions) for selected Fermi levels corresponding to
    $p$-doped samples. The ZGNR has 95 zigzag lines (about $20$~nm
    width). For panels where two different vertical scales are
    present, i.e. (b) and (c), the scale on the left (right) is for
    undoped (doped) cases (arrows below the factors indicate the
    ordinate for which they apply). A damping constant of 20~meV was
    introduced.}
  \label{fig:CURR_FERMIS}
\end{figure}

In Fig.~\ref{fig:TPAfermis}, \ref{fig:tpaers_fermis}, and
\ref{fig:CURR_FERMIS} we present the nonzero $\xi^{abcd}_\rmC$,
$\xi^{abcd}_\rmR$, and $\eta^{abcd}$ coefficients for selected nonzero
Fermi levels. As was seen for $\xi^{ab}$, doping the ZGNR has the
effect of modifying the responses around their onset energy, either
due to the removal of some transitions, or due to the appearance of
new ones, which in the undoped system were forbidden because the
initial and final states were filled [e.g. ($-1,-2$) or
($-1,-3$)]. This shows that doping is an effective way of modifying
the carrier and current injection in ZGNR, where the most significant
changes are due to the removal of density of states at the edge bands.

We close this section by mentioning that we performed finite
  temperature calculations at room temperature; this was achieved by
implementing a temperature dependence of the Fermi factors through the
Fermi-Dirac distribution.  We found that the only significant change
is in that the onset energy of the coefficients $\xi^{ab}$,
$\xi^{abcd}_{C(R)}$, and $\eta^{abcd}$ are smaller.
However, the magnitudes of the coefficients at energies near the lower
onsets are several orders of magnitude smaller that the magnitudes of
the corresponding coefficients at zero temperature near their energy
onsets.


\section{Limits of the model \label{sec:estimates}}
The model employed in this work inherits the limits of applicability
of time-dependent perturbation theory, which is restricted to
situations of low electron-hole pair densities~\cite{haug04} (for high
injection densities a density matrix formalism could be employed to
study the dynamics).
The regime of validity of the perturbation treatment used here can be
estimated: we require the populated fraction of excited states
accessible to a typical Gaussian pulse to be small.
\subsection{Graphene sheet}
As a reference, we first consider monolayer graphene.  When the
electric fields of the optical beams are all aligned along $\xhat$,
the one- and two-photon injection coefficients for a 2D graphene sheet
are \cite{Rioux2011} given by Eqs.~\eqref{eq:opa2Dxx} and~
\eqref{eq:tpa2Dxxxx}.  %
For each of $\xi^{xx}_{2D}$ and $\xi^{xxxx}_{2D}$, we set the number of
carriers injected per unit area to be less than the number of states 
per unit area accessible to the optical beam.  Then taking the beam
intensity as $I(\omega) = 2 \epsilon_0 c \abs{E(\omega)}^2$, we arrive
to
\begin{eqnarray}
  \label{eq:18a}
  I(2\omega) &<& 
\frac
{\epsilon_0 c \,\alpha\omega}
{2\pi v_F^2 (\Delta t)^2 \xi_{\mathrm{2D}}^{xx}(2\omega)},
\\
  \label{eq:18b}
I^{2}(\omega) &<&
\frac %
{ (2\epsilon_0 c)^2 \alpha\omega} %
{2\pi v_F^2 (\Delta t)^2 \xi_{\mathrm{2D}}^{xxxx}(\omega)},
\end{eqnarray}
where $\alpha$ is the time-bandwidth product for the optical beam
(which we take as 0.44, typical for a Gaussian beam), $\Delta t$ is
the pulse-duration, and $v_F\approx 10^6$~m/s is graphene's Fermi
velocity.

\subsection{Zigzag nanoribbons}
The estimate for the nanoribbon case is similar to the graphene sheet,
aside from the fact that the areal ratios become length ratios, i.e.
for each one of OPA and TPA coefficients we set the number of carriers
injected per unit \textit{length} to be less than the number of states
per unit \textit{length} accessible to the optical beam, giving us
\begin{eqnarray}
  \label{eq:LimitingIntensity}
 I(2\omega)
&<&
\frac {
\epsilon_0 c 
\,
\alpha }
{
\pi
(\Delta t)^2
\xi^{xx} (2\omega) 
( |v_e| +|v_h|)
},\\
 I^{2}_{\rmC(\rmR)}(\omega) \label{eq:intensityLimit}
&<&
\frac {
(2\epsilon_0 c)^2 
\,
\alpha }
{
\pi
(\Delta t)^2
\xi^{xxxx}_{\rmC(\rmR)} (\omega) 
( |v_e| +|v_h|)
},
\end{eqnarray}
where $\alpha$ and $\Delta t$ where defined previously, $v_e$ is the
velocity of the injected electrons in the conduction band, given by
the matrix element $v_{nn}$, and $v_h$ is the velocity of the holes
injected in the valence band, given by
$v_{mm}$. Equation~\eqref{eq:intensityLimit} provides the expression for
the conventional ($\rmC$) and ERS processes ($\rmR$).

In order to compare the limiting intensities of our model for a
graphene   sheet and for ZGNR, we assume a
typical pulse duration of $220$~fs and beam wavelengths of $3.2\mu m$
and $1.6\mu m$ for the $\omega$ and $2\omega$
beams\cite{sun-norris2010}. Then we identify the states that
contribute at these two wavelengths, and find that, on
average, $|v_e|+|v_h|\approx v_F$. From Eqs.~\eqref{eq:18a}
and \eqref{eq:LimitingIntensity}, at $\lambda=1.6~\mu$m,
\begin{eqnarray}
  \label{eq:ratio}
\frac %
{I^{\mathrm{Graphene}}(2\omega)} %
{I^{\mathrm{Ribbons}}(2\omega)} %
= %
\frac %
{\omega\, \xi^{xx}(2\omega)}
{2 v_F\, \xi_{\mathrm{2D}}^{xx}(2\omega)}
\approx 2.6,
\end{eqnarray}
and from Eqs.~\eqref{eq:18b} and \eqref{eq:intensityLimit}, at $\lambda=3.2~\mu$m,
\begin{eqnarray}
  \label{eq:ratioOmega}
\frac
{  I^{\mathrm{Graphene}}(\omega)} 
{ I^{\mathrm{Ribbons}}(\omega)} = %
\sqrt{%
\frac
{\omega\, \xi^{xxxx}_{\rmC}(\omega)}
{2 v_F\, \xi_{\mathrm{2D}}^{xxxx}(\omega)}
} 
\approx 1.6.
\end{eqnarray}
Equations~\eqref{eq:ratio} and \eqref{eq:ratioOmega} indicate that the
limiting intensities of our model are similar for a graphene sheet and
for a ZGNR, within an order of magnitude.

We find that, under the assumptions made in this section, the
estimated limit for the beam intensities at $\omega$ in the ZGNR and
the 2D graphene are about two orders of magnitude below the
intensities used in some experiments \cite{sun-norris2010} on 2D
graphene, where coherent current injection was observed. %
Due to relaxation processes, of course, the number of allowed
carrier excitations below saturation is expected to be higher than our
estimates, leading to larger values of the beam intensities for which
a perturbation approach would be valid. 
Based on the estimates in Eqs.~\eqref{eq:ratio} and \eqref{eq:ratioOmega},
if relaxation processes affect the ribbon samples as effectively as they
do for 2D samples, we can expect coherent control in ZGNR to be
observable at the higher intensities used in 2D graphene experiments.
%
%
%
%
%

\section{Summary and Discussion \label{sec:discussion}}
We have calculated the response coefficients for one- and two-photon
charge injection and the two-color current injection in a graphene
zigzag nanoribbon; we use the semi-empirical $\bm{k}\cdot\bm{p}$
method to describe the electron wavefunctions by smooth envelope
functions.

The only nonzero one-photon injection coefficients correspond to the
case of all-$x$ or all-$y$ aligned fields, i.e., $\xi^{xx}$ and
$\xi^{yy}$. These two coefficients possess a rich structure of 
 divergences, caused by divergences of the
joint-density-of-states originating from the infinite set of parabolic
bands present in the zigzag nanoribbon. These two coefficients have
distinct  selection rules for the allowed transitions.

The two-photon carrier injection coefficients drop off as the fifth
power of the photon energy at large photon energies, as they do for
monolayer graphene.
Moreover, these coefficients possess two classes of divergencies. One
corresponds to the joint-density-of-states divergences associated with
the parabolic bands. The second class corresponds to divergences
arising from resonant conditions, when the two-photon absorption
processes arise from sequential one-photon absorption processes
between real states. %
In our calculation here we broadened these resonances
phenomenologically, but a more sophisticated treatment of these
resonantly enhanced transitions is an outstanding problem on which we
hope this work will encourage further study.
The onset of the signals is determined by the minimum energy band gap and the
selection rules for these coefficients.

We calculated the electron and hole contributions to the conventional
and the stimulated electronic Raman scattering (ERS) current injection
processes, finding that the only nonzero components are associated
with current injected along the length of the nanoribbon, as
expected. The behavior of these coefficients as a function of the
photon energy follows the behavior of 2D graphene
[$\sim (\hbar\omega)^{-3}$] at large photon energies, aside of the
resonances present in the ribbons. We have also calculated the
so-called swarm velocity of the injected electrons, which inherits a
rich structure as a function of the photon energy due to the details
of the structure of the injection coefficients. All these calculations
were presented for a system at zero Fermi level and zero temperature.
However, we also carried finite temperature calculations and found
that, within this model, finite temperatures only account for changes
at the onset of the signals, which are several orders of magnitude
smaller than the nominal values at zero temperature.

Lower bound estimates on the permissible incident intensities for
which the calculations here can be valid were presented. They are
similar to those of monolayer graphene, where coherent current
injection has been observed at much higher intensities than these
simple estimates, which do not take into account the relaxation
effects in the excited populations. Thus experiments to demonstrate
coherent current injection in ZGNR seem to us to be in order.

For experiments contemplated for ribbons of different width than those
studied here, it is important to note that simple scaling arguments
show that the wider the ribbon, the stronger the confinement of the
energy bands. As shown in this work, at low photon energies, the
band gap follows a linear relation with respect to the inverse of the
ribbon width.  Consequently, increasing the width of the ribbon
decreases the energy band gap between any pair of bands.  This in turn
shifts the onset energy of the response coefficients towards zero
energy and increases the number of JDOS divergences per photon
energy. For instance, the onset of the response coefficients
  when light is polarized along the length of the ribbon is determined
  by the bangap between bands $(1,-2)$ (see Fig.~\ref{fig:bands_K}).
  For such pair of bands, a linear fit shows that the band gap depends
  on the ribbon width $W$ as $E^{\mathrm{gap}}_{1,-2}\approx a W^{-1}$
  with $a=2.98~e\mathrm{V}\cdot\mathrm{nm}$.
Besides altering the onset energy of the responses, a larger width
also leads to a larger magnitude of the injection coefficients, larger
than would be expected simply on the basis of the increase in
material; e.g., a width increase of about $15\%$ doubles the size of
$\eta^{xxxx}$.

As the outstanding signature of the zigzag nanoribbons are the
strongly localized edge states, we have identified their contribution
to the carrier- and current-injection processes.
In all cases the edge states always participate in the onset of the
signals. This lead us to consider a second scenario to study these
localized states: given that the dispersion relations of these states
are flattened towards zero energy for certain regions in $k$-space, we
re-visited our calculations considering doped scenarios. We found that
that even small doping levels allow for significant changes around the
onset energy of the signals. This is because the large joint-density
of states present between the edge states is diminished with nonzero
chemical potentials. Due to the relative ease of doping  graphene
systems, the present work shows that zigzag nanoribbons offer an
excellent opportunity to investigate scenarios in which electrical
currents can be generated and controlled optically. %
While more sophisticated treatments of the electron states and the
inclusion of electron-electron
interaction~\cite{Yang_Cohen_Louie_2008, lado2015} will undoubtedly
add to the richness of the injection processes, we hope that the
description given here will motivate all-optical current injection
experiments.
Although coherent control has been studied and observed on graphene
sheets, zigzag graphene nanoribbons have the advantage of having
optical responses that depend strongly on the geometry and width of
the ribbon. Moreover, as shown in the literature, the localized states
present in these ribbons are highly sensitive to external fields,
doping and functionalization. All these characteristics endow graphene
zigzag ribbons with a richness absent in simpler graphene sheets.

\section{Acknowledgments}

C.S. acknowledges partial support from CONACYT (Mexico) and 
Rodrigo~A.~Muniz for useful discussions.  J.L.C. acknowledges the
support from EU-FET grant GRAPHENICS (618086), the ERC-FP7/2007-2013
grant 336940, and the FWO-Vlaanderen project G.A002.13N. 
J.E.S. and C.S. acknowledge support from the Natural Sciences and
Engineering Research Council of Canada (NSERC). All the authors
acknowledge A.~Paramekanti and M.~Killi for drawing our attention to this
problem. 

\clearpage

\onecolumngrid
 \appendix

\section{Velocity matrix elements\label{appendixVME}}
%
%
\begin{table*} [ht]
  \centering
  \caption{Velocity matrix elements at the Dirac point    $\bm{\mathrm{K}}$. %
    At a given $\kappa_x$, any of these matrix elements are purely real or
    purely imaginary (which is explicitly indicated by the presence
    (absence) of the imaginary unit $i$).  The corresponding expressions at  the other Dirac
      point    $\bm{\mathrm{K}'}$ are identical, except that the
      $\bm{\hat{x}}-$components of the matrix
      elements flip sign; the $\bm{\hat{y}}$-components of the  matrix elements  remain unchanged.
    The range of validity for this expressions is given in the
    third column. \\  \label{tab:vmetable}}
  \begin{tabular} { c  l  l  l }
\hline
\hline
\hspace*{4mm}\textbf{Type}\hspace*{4mm} & \hfill\textbf{Expression} \hfill\phantom{.} & \textbf{Conditions} \\
\hline &&\\
$_n$Conf &
$
v^x_{nm} (\kappa_x) = 
-4  v_F  \left(    \signConfined_{m}+\signConfined_n \right) %
\Aconf_n  \Aconf_{m}  %
\left [  \frac{\calKconf_{m} \sin (\calKconf_n   \Weff)  - \calKconf_n \sin (\calKconf_{m} \Weff ) }  {(\calKconf_{m})^2-(\calKconf_n)^2}  %
\right] 
$ &
\multirow{1}{*}{$ |n|\ge 2, |m|\ge 2, \forall \kappax$, } or\\ 
$\updownarrow$
&&\multirow{1} {*}{$ |n|\ge 2, |m|= 1,  \kappax<\Weff^{-1}$, or } \\ 
$_m$Conf & 
$
v^y_{nm} (\kappa_x) = 
-i\; 4  v_F  \left(    \signConfined_{m}  -  \signConfined_n\right) %
\Aconf_n  \Aconf_{m}  %
 \left [  \frac{\calKconf_{m} \sin (\calKconf_n   \Weff)  - \calKconf_n \sin (\calKconf_{m} \Weff ) }  {(\calKconf_{m})^2-(\calKconf_n)^2}  %
\right] 
$ &
\multirow{1}{*}{$ |n|=1 , |m|\ge 2,  \kappax<\Weff^{-1}$ } \\[2ex] 
\hline &&\\
$_n$Edge &
$
v^x_{nm} (\kappa_x)  = 
-4  v_F   \left( \signEdge_{m} + \signEdge_n \right) %
\AEdge_n  \AEdge_{m}  %
\left [  \frac{\calKedge_n \sinh (\calKedge_{m}   \Weff)  - \calKedge_{m} \sinh (\calKedge_n \Weff ) }  {(\calKedge_{m})^2-(\calKedge_n)^2}  %
\right]
$ &
\multirow{3}{*}{$ |n|\ge 1, |m|\ge 1, \kappax>\Weff^{-1}$ } \\ 
$\updownarrow$ &  & \\ 
$_m$Edge &$
v^y_{nm} (\kappa_x) =
- i \; 4 v_F    \left(   \signEdge_{m}  -   \signEdge_n\right) %
\AEdge_n  \AEdge_{m}  %
\left [  \frac{\calKedge_n \sinh (\calKedge_{m}   \Weff)  - z_{m} \sinh (\calKedge_n \Weff ) }  {(\calKedge_{m})^2-(\calKedge_n)^2}  %
\right]$ &\\[2ex]
\hline &&\\
$_n$Conf &
$
v^x_{nm} (\kappa_x) =
 \phantom{-}
i 4  v_F  \left(   \signEdge_{m} + \signConfined_n \right) %
\Aconf_n \; \AEdge_{m}
\left[  %
  \frac{\calKconf_n \sinh (\calKedge_{m}   \Weff)  - \calKedge_{m} \sin (\calKconf_n \Weff ) }  {(\calKedge_{m})^2+(\calKconf_n)^2}  %
  \right]
$ &
\multirow{3}{*}{$ |n|\ge 2, |m|= 1, \kappax>\Weff^{-1}$ } \\ 
$\updownarrow$ &  &\\ 
$_m$Edge &
$v^y_{nm} (\kappa_x) =
- 4 v_F 
 \left(
  \signEdge_{m}-\signConfined_n \right) %
\Aconf_n \; \AEdge_{m} 
\left[  %
  \frac{\calKconf_n \sinh (\calKedge_{m}   \Weff)  - \calKedge_{m} \sin (\calKconf_n \Weff ) }  {(\calKedge_{m})^2+(\calKconf_n)^2}  %
\right]$
&\\[2ex]
\hline &&\\
$_n$Edge & & \\
$\updownarrow$ &  $\left(\text{Conf}\leftrightarrow\text{Edge}\right)^\dagger$  &
$ |n|= 1, |m|\ge2, \kappax>\Weff^{-1}$     \\
$_m$Conf & & \\
\hline
\hline
  \end{tabular}
\end{table*}
%
%



\twocolumngrid
\bibliographystyle{unsrt} 
\bibliography{./references}

\begin{thebibliography}{10}

\bibitem{Ogawa_optics_low_dimensions}
T.Ogawa and Y.Kanemitsu.
\newblock {\em Optical Properties of Low–Dimensional Materials}.
\newblock World Scientific, 1996.

\bibitem{Nakada_edges_early}
Kyoko Nakada, Mitsutaka Fujita, Gene Dresselhaus, and Mildred~S. Dresselhaus.
\newblock Edge state in graphene ribbons: Nanometer size effect and edge shape
  dependence.
\newblock {\em Phys. Rev. B}, 54:17954--17961, Dec 1996.

\bibitem{BreyFertig}
L.~Brey and H.~A. Fertig.
\newblock Electronic states of graphene nanoribbons studied with the {D}irac
  equation.
\newblock {\em Phys. Rev. B}, 73:235411, Jun 2006.

\bibitem{marconciniKDP}
P.~Marconcini and M.~Macucci.
\newblock The $\bm{k}\cdot\bm{p}$ method and its application to graphene,
  carbon nanotubes and graphene nanoribbons: the {D}irac equation.
\newblock {\em La Rivista del Nuovo Cimento}, pages 489--584, 2011.

\bibitem{Lee_functionalization}
Hoonkyung Lee, Marvin~L. Cohen, and Steven~G. Louie.
\newblock Selective functionalization of halogens on zigzag graphene
  nanoribbons: A route to the separation of zigzag graphene nanoribbons.
\newblock {\em Applied Physics Letters}, 97(23):233101--1, 2010.

\bibitem{daRocha_functionalization}
Gomes da~Rocha, Clayborne, Koskinen, and Hakkinen.
\newblock Optical and electronic properties of graphene nanoribbons upon
  adsorption of ligand-protected aluminum clusters.
\newblock {\em Phys. Chem. Chem. Phys.}, 16:3558--3565, 2014.

\bibitem{Yang_Cohen_Louie_2008}
Li~Yang, Marvin~L. Cohen, and Steven~G. Louie.
\newblock Magnetic edge-state excitons in zigzag graphene nanoribbons.
\newblock {\em Phys. Rev. Lett.}, 101:186401, Oct 2008.

\bibitem{Kunstmann_Stability}
Jens Kunstmann, Cem \"Ozdo\ifmmode~\breve{g}\else \u{g}\fi{}an, Alexander
  Quandt, and Holger Fehske.
\newblock Stability of edge states and edge magnetism in graphene nanoribbons.
\newblock {\em Phys. Rev. B}, 83:045414, Jan 2011.

\bibitem{Bellec_edges_artificial_graphene}
M.~Bellec, U.~Kuhl, G.~Montambaux, and F.~Mortessagne.
\newblock Manipulation of edge states in microwave artificial graphene.
\newblock {\em New J. Phys.}, 16:113023, 2014.

\bibitem{Delplace_edges_zak_phase}
P.~Delplace, D.~Ullmo, and G.~Montambaux.
\newblock Zak phase and the existence of edge states in graphene.
\newblock {\em Phys. Rev. B}, 84:195452, Nov 2011.

\bibitem{Yazyev_applications}
Oleg~V. Yazyev.
\newblock A guide to the design of electronic properties of graphene
  nanoribbons.
\newblock {\em Accounts of Chemical Research}, 46(10):2319--2328, 2013.
\newblock PMID: 23282074.

\bibitem{luck2015}
J~M Luck and Y~Avishai.
\newblock Unusual electronic properties of clean and disordered zigzag graphene
  nanoribbons.
\newblock {\em Journal of Physics: Condensed Matter}, 27(2):025301, 2015.

\bibitem{Fujita_EELS_2015}
N~Fujita, P~J Hasnip, M~I~J Probert, and J~Yuan.
\newblock Theoretical study of core-loss electron energy-loss spectroscopy at
  graphene nanoribbon edges.
\newblock {\em Journal of Physics: Condensed Matter}, 27(30):305301, 2015.

\bibitem{pelc-brey}
Marta Pelc, Eric~Suárez Morell, Luis Brey, and Leonor Chico.
\newblock Electronic conductance of twisted bilayer nanoribbon flakes.
\newblock {\em The Journal of Physical Chemistry C}, 119(18):10076--10084,
  2015.

\bibitem{gonsalbez-spin-filtered}
D.~Gos\'albez-Mart\'inez, D.~Soriano, J.J. Palacios, and
  J.~Fern\'andez-Rossier.
\newblock Spin-filtered edge states in graphene.
\newblock {\em Solid State Communications}, 152(15):1469 -- 1476, 2012.
\newblock Exploring Graphene, Recent Research Advances.

\bibitem{Stegmann}
Thomas Stegmann and Axel Lorke.
\newblock Edge magnetotransport in graphene: A combined analytical and
  numerical study.
\newblock {\em Annalen der Physik}, 527(9-10):723--736, 2015.

\bibitem{lado2015}
J.L. Lado, N.~García-Martínez, and J.~Fernández-Rossier.
\newblock Edge states in graphene-like systems.
\newblock {\em Synthetic Metals}, 210, Part A:56 -- 67, 2015.
\newblock Reviews of Current Advances in Graphene Science and Technology.

\bibitem{Hsu-Reich_selection_rules}
Han Hsu and L.~E. Reichl.
\newblock Selection rule for the optical absorption of graphene nanoribbons.
\newblock {\em Phys. Rev. B}, 76:045418, Jul 2007.

\bibitem{Sasaki_optical_transitions}
Ken-ichi Sasaki, Keiko Kato, Yasuhiro Tokura, Katsuya Oguri, and Tetsuomi
  Sogawa.
\newblock Theory of optical transitions in graphene nanoribbons.
\newblock {\em Phys. Rev. B}, 84:085458, Aug 2011.

\bibitem{Yamamoto2006_TB_optics_flakes}
Takahiro Yamamoto, Tomoyuki Noguchi, and Kazuyuki Watanabe.
\newblock Edge-state signature in optical absorption of nanographenes:
  Tight-binding method and time-dependent density functional theory
  calculations.
\newblock {\em Phys. Rev. B}, 74:121409, Sep 2006.

\bibitem{Berahman_optics_chiral}
M.~Berahman, M.~Asad, M.~Sanaee, and M.H. Sheikhi.
\newblock Optical properties of chiral graphene nanoribbons: a first principle
  study.
\newblock {\em Optical and Quantum Electronics}, 47(10):3289--3300, 2015.

\bibitem{zhu_optical_conduc}
Wen-Huan Zhu, Guo-Hui Ding, and Bing Dong.
\newblock The enhanced optical conductivity for zigzag-edge graphene
  nanoribbons with applied gate voltage.
\newblock {\em Applied Physics Letters}, 100(10):103101--1, 2012.

\bibitem{Prezzi_2008_optics}
Deborah Prezzi, Daniele Varsano, Alice Ruini, Andrea Marini, and Elisa
  Molinari.
\newblock Optical properties of graphene nanoribbons: The role of many-body
  effects.
\newblock {\em Phys. Rev. B}, 77:041404, Jan 2008.

\bibitem{ManykinAfanas}
E.A. Manykin and A.M. Afanas'ev.
\newblock On one possibility of making a medium transparent by multiquantum
  resonance.
\newblock {\em J.Exptl. Theor. Phys.}, 25(2):828--830, November 1967.
\newblock Original in Russian: ZhETF \textbf{52}, No.~5, p.~1246-1249 (1967).

\bibitem{Manykin}
E.A. Manykin.
\newblock Quantum interference and coherent control.
\newblock {\em Laser Phys.}, 11:60, 2001.

\bibitem{ccontrolDrielSipe2001}
H.M. van Driel and J.E. Sipe.
\newblock {\em Ultrafast Phenomena in Semiconductors}, chapter 5: Coherent
  Control of Photocurrents in Semiconductors, pages 261--306.
\newblock Springer, 2001.

\bibitem{sun-norris2010}
Dong Sun, Charles Divin, Julien Rioux, John~E. Sipe, Claire Berger, Walt~A.
  de~Heer, Phillip~N. First, and Theodore~B. Norris.
\newblock Coherent control of ballistic photocurrents in multilayer epitaxial
  graphene using quantum interference.
\newblock {\em Nano Letters}, 10(4):1293--1296, 2010.
\newblock PMID: 20210362.

\bibitem{Rioux2011}
J.~Rioux, Guido Burkard, and J.~E. Sipe.
\newblock Current injection by coherent one- and two-photon excitation in
  graphene and its bilayer.
\newblock {\em Phys. Rev. B}, 83:195406, May 2011.

\bibitem{fregoso}
Benjamin~M Fregoso and Sinisa Coh.
\newblock Intrinsic surface dipole in topological insulators.
\newblock {\em Journal of Physics: Condensed Matter}, 27(42):422001, 2015.

\bibitem{kiran2012}
Kiran~M. Rao and J.~E. Sipe.
\newblock Coherent photocurrent control in graphene in a magnetic field.
\newblock {\em Phys. Rev. B}, 86:115427, Sep 2012.

\bibitem{Muniz_SpinTopos}
Rodrigo~A. Muniz and J.~E. Sipe.
\newblock Coherent control of optical injection of spin and currents in
  topological insulators.
\newblock {\em Phys. Rev. B}, 89:205113, May 2014.

\bibitem{Muniz_ValleyCurrent15}
Rodrigo~A. Muniz and J.~E. Sipe.
\newblock All-optical injection of charge, spin, and valley currents in
  monolayer transition-metal dichalcogenides.
\newblock {\em Phys. Rev. B}, 91:085404, Feb 2015.

\bibitem{CastroNetoReview2009}
A.~H. Castro~Neto, F.~Guinea, N.~M.~R. Peres, K.~S. Novoselov, and A.~K. Geim.
\newblock The electronic properties of graphene.
\newblock {\em Rev. Mod. Phys.}, 81:109--162, Jan 2009.

\bibitem{DasSarmaReview2011}
S.~Das~Sarma, Shaffique Adam, E.~H. Hwang, and Enrico Rossi.
\newblock Electronic transport in two-dimensional graphene.
\newblock {\em Rev. Mod. Phys.}, 83:407--470, May 2011.

\bibitem{Note1}
The graphene's honeycomb lattice is composed by two distinct triangular
  lattices, A and B. On each sub-lattice all atoms are equivalent.

\bibitem{RiouxERS}
J.~Rioux, J.~E. Sipe, and Guido Burkard.
\newblock Interference of stimulated electronic raman scattering and linear
  absorption in coherent control.
\newblock {\em Phys. Rev. B}, 90:115424, Sep 2014.

\bibitem{ccontrolDrielSipe2005}
H.M. van Driel and J.E. Sipe.
\newblock Coherent control: Applications in semiconductors.
\newblock In Robert~D. Guenther, editor, {\em Encyclopedia of Modern Optics},
  pages 137--143. Elsevier, Oxford, 2005.

\bibitem{dropoff}
At large photon energies, the two-photon absorption coefficients for zigzag
  nanoribbons drop off with the fifth power of the photon energy, as they do
  for a monolayer of graphene.

\bibitem{interpNote}
We handle Dirac delta integrals of the form $I(\omega) = \int dk
  F_{nmk}\delta(\omega-\omega_{nmk})$ by doing an interpolation of the
  integrand, such that\\ $ I(\omega) = \sum_{i=0}^{N-1}
  \left[\frac{k_{i+1}-Q_i}{k_{i+1}-k_i}\frac{F_{nmk_i}}{|\Delta_{nmk_i}|} +
  \frac{Q_i-k_{i}}{k_{i+1}-k_i}\frac{F_{nmk_{i+1}}}{|\Delta_{nmk_{i+1}}|}\right]$\\
  $\times\theta(Q_i-k_i) \theta(k_{i+1}-Q_i)$, where $Q_i$ is the interpolated
  $k$-point that satisfies the Dirac delta,
  $Q_i=\frac{k_{i+1}(\omega-\omega_{nmk_i})+k_i(\omega_{nmk_{i+1}}-\omega)}{\omega_{nmk_{i+1}}-\omega_{nmk_i}}$,
  and $\theta(k)$ is the unit step function. This interpolation scheme requires
  convergence on a single parameter, the number of $k$ points. More simple
  numerical treatments of the Dirac delta integrals with broadening functions
  (Lorentzian or Gaussian functions) require a larger number of $k$-points to
  reach convergence.

\bibitem{haug04}
Hartmut Haug and Stephan~W. Koch.
\newblock {\em Quantum Theory of the Optical and Electronic Properties of
  Semiconductors}.
\newblock World Scientific Publishing Company, 4 edition, 2004.

\end{thebibliography}
\nocite{*}

\end{document}